\documentclass[pdftex,twocolumn,nofootinbib,preprintnumbers,superscriptaddress]{revtex4-1}

\usepackage{amsmath}
\usepackage{amssymb}
\usepackage{listings}
\usepackage{hyperref}
\usepackage{dsfont}
\usepackage{slashed}
\usepackage{color}
\usepackage{array}
\usepackage{graphicx}
\usepackage{epstopdf}
\usepackage{subfigure}
\usepackage{epsfig}
\usepackage{latexsym}
\usepackage{multirow}
\usepackage{mciteplus}
\usepackage{bm}

\usepackage{hhline}
\usepackage{lineno}

\newcommand*{\LessApprox}{\smallrel\lessapprox}
\newcommand*{\GtrSim}{\smallrel\gtrsim}

\makeatletter
\newcommand*{\smallrel}[2][.8]{%
  \mathrel{\mathpalette{\smallrel@{#1}}{#2}}%
}
\newcommand*{\smallrel@}[3]{%
  \sbox0{$#2\vcenter{}$}%
  \dimen@=\ht0 %
  \raise\dimen@\hbox{%
    \scalebox{#1}{%
      \raise-\dimen@\hbox{$#2#3\m@th$}%
    }%
  }%
}
\makeatother

\begin{document}


\title{Measurement of the Gerasimov-Drell-Hearn integrand for
proton and deuteron
  from 200 to 1400 MeV}
%


%
\newcommand{\mainz}{Institut f\"ur Kernphysik, University of Mainz, D-55099 Mainz, Germany}
\newcommand{\basel} {Department f\"ur Physik, Universit\"at Basel, CH-4056 Basel, Switzerland}
\newcommand{\bonn}{Helmholtz-Institut f\"ur Strahlen- und Kernphysik, Universit\"at Bonn, D-53115 Bonn, Germany}
\newcommand{\infn}{INFN Sezione di Pavia, I-27100 Pavia, Italy}

\newcommand{\glasgow}{SUPA School of Physics and Astronomy, University of Glasgow, Glasgow G12 8QQ, United Kingdom}
\newcommand{\york}{Department of Physics, University of York, Heslington, York, Y010 5DD, UK}
\newcommand{\dubna}{Joint Institute for Nuclear Research, 141980 Dubna, Russia}
\newcommand{\unipv}{Dipartimento di Fisica, Università di Pavia, Pavia; Italy}
\newcommand{\GWU}{The George Washington University, Washington, DC 20052-0001, USA}
\newcommand{\novos}{Budker Institute of Nuclear Physics, 630090 Novosibirsk, Russia}
\newcommand{\INR}{Institute for Nuclear Research, 125047 Moscow, Russia}
\newcommand{\mta}{Mount Allison University, Sackville, New Brunswick E4L 1E6, Canada}
\newcommand{\Kent}{Kent State University, Kent, Ohio 44242-0001, USA}
\newcommand{\Amherst}{University of Massachusetts, Amherst, Massachusetts 01003, USA}
\newcommand{\Zagreb}{Rudjer Boskovic Institute, HR-10000 Zagreb, Croatia}
\newcommand{\Regina}{University of Regina, Regina, Saskatchewan S4S 0A2, Canada}
\newcommand{\bass}{Kitzb\"uhel Centre for Physics, Kitzb\"uhel, Austria}
\newcommand{\bast}{Marian Smoluchowski Institute of Physics, Jagiellonian University , Krakow, Poland}
\author{P.~Pedroni}\footnote{Corresponding author; email:pedroni@pv.infn.it}
\affiliation{\infn}
\author{F.~Afzal}\affiliation{\bonn}
\author{S.~Abt}\affiliation{\basel}
\author{P.~Achenbach}\affiliation{\mainz}
\author{J.R.M.~Annand}\affiliation{\glasgow}
\author{H.J.~Arends}\affiliation{\mainz}
\author{S.D.~Bass}\affiliation{\bass}\affiliation{\bast}
\author{M.~Biroth}\affiliation{\mainz}
\author{R.~Beck}\affiliation{\bonn}
\author{N.~Borisov}\email{Deceased}
\affiliation{\dubna}
\author{A.~Braghieri}\affiliation{\infn}
\author{W.J.~Briscoe}\affiliation{\GWU}
\author{F.~Cividini}\affiliation{\mainz}
\author{C.~Collicott}\affiliation{\mainz}
\author{A.~S.~Dolzhikov}\affiliation{\dubna}
\author{E.~Downie}\affiliation{\GWU}
\author{S.~Fegan}\affiliation{\york}
\author{A.~Fix}\affiliation{\novos}
\author{D.~Ghosal}\affiliation{\basel}
\author{I.~Gorodnov}\affiliation{\dubna}
\author{W.~Gradl}\affiliation{\mainz}
\author{G.~Gurevich}\email{Deceased}
\affiliation{\INR}
\author{L.~Heijkenskj\"old}\affiliation{\mainz}
\author{D.~Hornidge}\affiliation{\mta}
\author{G.M.~Huber}\affiliation{\Regina}
\author{V.L.~Kashevarov}\affiliation{\mainz}
\author{S.J.D.~Kay}\affiliation{\york}
\author{M.~Korolija}\affiliation{\Zagreb}
\author{B.~Krusche}\email{Deceased}
\affiliation{\basel}
\author{A.~Lazarev}\affiliation{\dubna}
\author{K.~Livingston}\affiliation{\glasgow}
\author{S.~Lutterer}\affiliation{\basel}
\author{I.J.D.~MacGregor}\affiliation{\glasgow}
\author{D.M.~Manley}\affiliation{\Kent}
\author{P.P.~Martel}\affiliation{\mainz}
\author{R.~Miskimen}\affiliation{\Amherst}
\author{M.~Mocanu}\affiliation{\york}
\author{E.~Mornacchi} \affiliation{\mainz}
\author{C.~Mullen}\affiliation{\glasgow}
\author{A.~Neganov}\affiliation{\dubna}
\author{A.~Neiser}\affiliation{\mainz}
\author{M.~Oberle}\affiliation{\basel}
\author{M.~Ostrick}\affiliation{\mainz}
\author{P.B.~Otte}\affiliation{\mainz}
\author{D.~Paudyal}\affiliation{\Regina}
\author{A.~Powell}\affiliation{\glasgow}
\author{T.~Rostomyan}\altaffiliation{Now at Paul Scherrer Institute (PSI), CH-5232 Villigen PSI, Switzerland.}
\affiliation{\basel}
\author{V.~Sokhoyan}\affiliation{\mainz}
\author{K.~Spieker}\affiliation{\bonn}
\author{O.~Steffen}\affiliation{\mainz}
\author{I.I.~Strakovsky}\affiliation{\GWU}
\author{T.~Strub}\affiliation{\basel}
\author{M.~Thiel}\affiliation{\mainz}
\author{A.~Thomas}\affiliation{\mainz}
\author{Yu.A.~Usov}\affiliation{\dubna}
\author{S.~Wagner}\affiliation{\mainz}
\author{D.P.~Watts}\affiliation{\york}
\author{D.~Werthm\"uller}\altaffiliation{Now at Paul Scherrer Institute (PSI), CH-5232 Villigen PSI, Switzerland.}
\affiliation{\york}
\author{J.~Wettig}\affiliation{\mainz}
\author{L.~Witthauer}\affiliation{\basel}
\author{M.~Wolfes}\affiliation{\mainz}
\author{N.~Zachariou}\affiliation{\york}

\collaboration{A2 Collaboration at MAMI}
%
\date{}
%
%
\begin{abstract}
  New data  for the total inclusive helicity-dependent cross section
  for the proton and deuteron were obtained in the photon energy
  interval 200$-$1400~MeV. 
  The experiment was performed at the A2 tagged-photon facility of the
  Mainz Microtron (MAMI) using a circularly polarized photon beam and
  longitudinally polarized proton and deuteron targets.
  The reaction products were detected using the large-acceptance Crystal
  Ball/TAPS calorimeter, which covers 97\% of the full solid angle.
  These new results, obtained with fine energy binning,
  significantly expand both the quantity and the quality of the available
  data for these observables and enable a detailed comparison with
  state-of-the-art theoretical calculations.
  From the combination of the results for the deuteron and the proton,
  important information could also be extracted for the neutron.
  Based on these data, and using existing models to evaluate the missing
  contributions from unmeasured photon energy regions, the validity of the
  Gerasimov-Drell-Hearn (GDH) sum rule has been verified for the proton,
  the neutron, and the deuteron.
  These new data provide a precise experimental benchmark for theoretical
  models used to study nucleons, both in their free state and when embedded
  in the nuclear medium.

\end{abstract}

\maketitle

\section{Introduction}

The Gerasimov-Drell-Hearn (GDH) sum rule establishes a fundamental
link between the total absorption cross section of circularly polarized
photons on longitudinally polarized nucleons and nuclei and
their static properties~\cite{Gera,Drel}.
The two
relative spin configurations, parallel or antiparallel, determine
the two absorption cross sections $\sigma_{P}$ and
$\sigma_{A}$.
In the target rest frame,
the integral over the photon energy E$_\gamma$ of the
difference of these two cross sections, weighted by the inverse
of E$_\gamma$, is related to the mass $M$, spin $S$, and anomalous
magnetic moment $\kappa$ of the nucleon (nucleus) as follows:
\begin{equation}\label{eq1}
I =
\int_{\nu_0}^\infty \Delta\sigma\,{\hbox{d}\hbox{E}_\gamma \over{\hbox{E}_\gamma} }
= \frac{4\pi^2 \alpha}{M^2} S\kappa^2
\end{equation}
where $\Delta\sigma= (\sigma_{P} - \sigma_{A})$,
$\nu_0$ is the minimal photon energy that can be absorbed by the target
and $\alpha$ is the fine-structure constant.

This key sum-rule, formulated in the 1960's, rests upon basic
physics principles (Lorentz-invariance, gauge invariance,
unitarity) and an unsubtracted dispersion relation applied to the
forward Compton amplitude (for comprehensive reviews of the GDH sum rule
and related integrals, see~\cite{drecrev,helb,igor}).
It implies that a non-zero anomalous magnetic moment $\kappa$ is
directly connected to the excitation spectrum of the nucleon (nucleus)
and, therefore, to its composite nature.

Experiments performed by the GDH collaboration both at Bonn (ELSA) and Mainz
(MAMI) have measured for the first time the GDH integrand for a
proton (deuteron) target through the range of incident photon energies
E$_\gamma = 0.2 - 0.8$~GeV (MAMI) and $0.7- 2.9\, (1.8)$~GeV
(ELSA)~\cite{ahr00,ahr01,dutz03,dutz04,dutz05,ahr06b,ahr09}.
The combination of these experimental results
with model-dependent estimates of 
the contributions of the unmeasured
photon energy intervals, yield an estimate that
is consistent with the GDH sum rule values of the proton ($205$~$\mu$b)
within the experimental uncertainties.

Recently, additional interest in the GDH sum rule has been stimulated by its connection to
understanding and describing the properties of nucleons bound in the nuclear
medium~\cite{Bass0,Bass1}.

In this context, both sides of Eq.~\ref{eq1} are expected to be modified.
This is because the nucleon magnetic moment is predicted to be enhanced
(see, for instance,~\cite{saito,Deur:2025inv}), and the contribution of the $\Delta(1232)$
resonance excitation, which constitutes the main contribution to the left-hand
side integral, will occur at slightly lower incident photon energies compared
to the free-nucleon case, due to nuclear binding energy effects.
Thus, measuring the GDH sum rule integral for different nuclear targets
constitutes an extremely sensitive tool for the quantitative experimental
study of
possible medium modifications, a fundamental problem in nuclear physics.

For these studies, it is crucial to have highly precise data for both the
free nucleon and the deuteron (the simplest nucleon bound state), particularly
in the first and second resonance regions, where nuclear effects are
most significant (see, for instance,~\cite{cividini} and references therein).
In both cases, microscopic theoretical  models can be used
to perform a precise isospin decomposition of the free reaction amplitudes.
In line with these objectives,
ongoing plans to study the convergence of the GDH integral for the
nucleon~\cite{jlabregge} and the helicity dependence of the total inclusive
photoabsorption cross section on light and medium nuclei
are currently under development at the JLab facility
(see, for instance, \cite{jlabx, afzalwp}).

In this paper, we present a precise measurement of the GDH integrand for both
the proton and the deuteron at incident photon energies from 
200~MeV up to approximately 1.4 GeV. This experiment was performed
at the tagged-photon facility of the MAMI electron accelerator in Mainz, Germany~\cite{mamic},
using the experimental setup of the A2 collaboration.
From the combination of the results obtained for the deuteron and the proton,
important information has also been deduced for the neutron.


\section{Experimental setup}\label{expapp}

The helicity-dependent data used for this analysis were collected in different
beamtime periods at the A2 tagged photon facility of the MAMI electron
accelerator in Mainz, Germany~\cite{mamic}.

\begin{figure*}%
\centering
\includegraphics[scale=0.27]{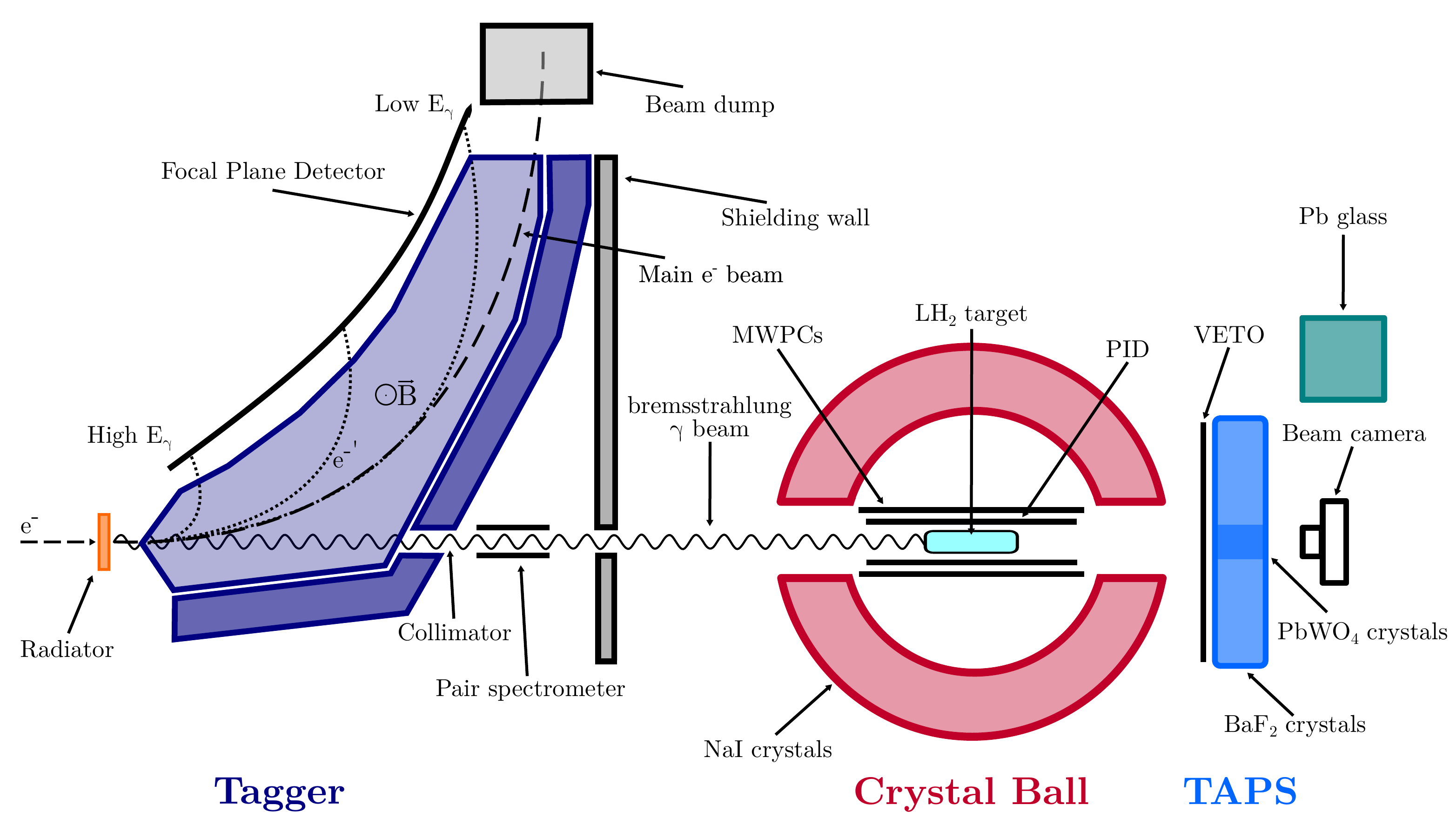}
\caption{Sketch of the experimental setup of the A2 Collaboration tagged photon facility,
  including photon tagging apparatus and detectors~\cite{edo}. The figure is not to scale.}
\label{fig_A2_setup}
\end{figure*}

Figure~\ref{fig_A2_setup} shows a sketch of the A2 experimental setup used for the measurement.
Since this setup has already been described in detail (see, for instance,
Refs.~\cite{A2:2013wkp,A2:2014pie,diet1,witt1,diet2,cividini} and
references therein), we will limit the discussion to those features relevant to the present experiment.


\subsection{The photon beam}

The circularly polarized photons used for this measurement were produced
using bremsstrahlung of the longitudinally polarized electrons
on an amorphous radiator from a longitudinally polarized electron beam.
Electron energies of both 450~MeV and 1557~MeV were used to collect the data presented here.

To avoid polarization dependent photon flux values,
the helicity of the electron beam was flipped at a rate of 1~Hz.
The electron polarization degree, $P_e$, was regularly
determined by Mott scattering close to the electron source~\cite{mott} 
and was found to be more than 80\% with a systematic uncertainty of $\pm 3\%$.
In addition, Moeller scattering at the bremsstrahlung
radiator was used as a secondary 
polarisation monitor.

%
The recoil electrons from the bremsstrahlung process were momentum-analyzed
using the Glasgow-Mainz spectrometer with an energy resolution,
depending on the electron beam energy, of
$\sim 1-4$~MeV, which
corresponds to the width of the detector channels~\cite{taggnew}. 
The resulting photon beam passed through a 2~mm-diameter lead collimator,
before reaching the target and detection apparatus.

The degree of the energy-dependent circular photon polarization, $P_{\odot}^{\gamma}$,
was determined using the Olsen and Maximon equation~\cite{olsen}:
\begin{equation*}
  \frac{P_{\odot}^{\gamma}}{P_e} = \frac{4x-x^2}{4-4x+3x^2} \ ,   
\end{equation*}
where $x=  E_e / E_\gamma$, with $E_e$ and $E_\gamma$ being the energies
of the electron and the
bremsstrahlung photon, respectively.     
 
The photon tagging efficiency was measured once a day using a Pb-Glass Cherenkov detector
in dedicated low flux runs.
During the standard data taking operation, fluctuations in the photon flux 
were monitored using a low-efficiency pair spectrometer 
located in the photon beamline after the collimator.
An absolute systematic uncertainty in the photon flux of 4\% was
estimated by comparing the data from these detectors, obtained under
a range of different experimental conditions.

\subsection{The target system}

The longitudinally polarized proton (deuteron) target used in this experiment
was the Mainz-Dubna Frozen Spin Target~\cite{Rohl,Thomas}.
The filling factor for the $\sim 2$-mm-diameter butanol
spheres contained in the 2-cm-long, 2-cm-diameter target container
was estimated to be 60\%, with a systematic uncertainty of 2\%~\cite{Rohl}.

The target material (butanol or deuterated butanol) was polarized using the
Dynamic Nuclear Polarization effect~\cite{Brad99},
which requires a high magnetic field (about 2.5~T) and a temperature of about 25~mK.
A small holding magnetic field of 0.6~T, which replaced the polarizing magnet
during the data taking phase, allowed
regular relaxation times of about 1000~h to be achieved.

The target polarization was measured with an NMR system
before and after each data taking period and then
exponentially interpolated at intermediate times.
Corrections to the calculated polarization values of the proton (butanol)
target were necessary due to ice formation on the NMR coils. These corrected
values were taken from Refs.~\cite{dilli,A2:2019bqm,farah}, which
independently analyzed the same proton data sets used in the present work.

Corrections were also applied to a portion of the deuteron (deuterated butanol)
data due to small inhomogeneities in the field of the polarizing magnets.
In this case, the correction values were taken from Refs.~\cite{cividini},
which independently analyzed the same deuteron data sets.

As in all previous references, a conservative systematic uncertainty of 10\%
was assigned to all evaluated target polarization values to account for
the effects of these corrections.

Some dedicated data runs with a carbon target were also performed to
determine the background contributions from the unpolarized carbon and oxygen
nuclei present in the target material. The carbon target was made from a
foam of the same density as the heavy nuclei in the butanol target. It was
fitted into the butanol's Teflon container to ensure identical geometry and
surrounding materials.

 
\subsection{The hadron detector} \label{sec-trigger}

{
The 
photon-induced reaction products 
were detected 
by using the CrystalBall (CB)~\cite{artcb} as the central calorimeter and
TAPS\footnote{TAPS is an acronym for Two-/Three-Arm Photon Spectrometer, since it was previously used as a stand-alone detector.}~\cite{taps1,taps2} as the forward calorimeter, which was installed  approximately 1.5~m downstream
of the CB center.}
%
The CB detector was placed around the target cell
and covered the full azimuthal ($\phi$) angle and a polar ($\theta$) angle range from 21$^{\circ}$ to 159$^{\circ}$.
It consisted of 672 NaI(Tl) crystals and had a $\sim 100\%$ detection efficiency
for photons coming from $\pi^0$ decay.
Inside the CB, starting from the center,
there were a Particle Identification Detector (PID),
consisting of a barrel of 24 plastic scintillators, followed by
two Multi-Wire Proportional Chambers (MWPCs).
The combination of these detectors provided  precise tracking
and identification of charged particles. 
TAPS was a hexagonal wall covering the polar angle forward region outside the CB acceptance,
$5^{\circ} < \theta < 20^{\circ}$, and was made of 
366 hexagonal BaF$_2$ and 72 PbWO$_{4}$ crystals.
In front of each TAPS crystal there was a 5-mm-thick
plastic scintillator (VETO)
that was used for charged particle identification.
The combination of the large acceptance CB and TAPS calorimeters covered
$\sim 97$\% of the full solid angle.

Two different triggers were used to collect the data presented here:

\begin{itemize}

  \item[(a)] For the 450~MeV beam energy setting, the total sum of pulse amplitudes from CB crystals 
    was required to exceed a hardware threshold corresponding to approximately $40$~MeV.
    This trigger setting was used for a
    subset of the polarized proton target runs analysed in the present work.

  \item[(b)] for the 1557~MeV beam energy setting, the total sum of pulse
    amplitudes from
    CB or TAPS crystals was required to be above a hardware threshold which varied
    between approximately~$40$ and $90$~MeV, depending on the specific data set.
    This threshold  energy setting was used for another subset of  the proton
    runs and for
    all the runs  with a polarized deuteron target.
   For this setting, an additional Cherenkov detector was placed in front of TAPS to online veto unwanted triggers in TAPS from e$^+$e$^-$ pairs produced by the photon beam.
   
\end{itemize}
   Independent analyses of all previous  beam time periods  have been presented in
   Refs.~\cite{cividini,dilli,A2:2019bqm,farah,Mornacchi:2024}.

\section{Data Analysis and unpolarized results} \label{sec:ana}

\subsection{Experimental method} \label{par:method}

Within the considered photon energy range, the photoabsorption process leads to 
many different multiparticle final states that are very difficult to
detect individually with full acceptance and efficiency.

To avoid the large systematic uncertainties associated with unobserved
final states, the total photoabsorption cross section must be measured
inclusively, as was done previously by the GDH
Collaboration~\cite{ahr01,dutz03,dutz04,dutz05,ahr06b,ahr09}.
The identification of individual processes is not necessary;
what is required is to detect at least one reaction product from all possible
hadronic final states with nearly complete acceptance
in terms of solid angle and efficiency.
This approach minimizes the corrections for detector inefficiencies and the
loss of events emitted into angular and momentum regions outside the
detector's coverage, thereby reducing model-dependent uncertainties.

Meeting these requirements necessitates both the reliable detection of
charged particles and high efficiency for detecting the neutral decay modes
of hadrons. The CB-TAPS detector,
which covers 97\% of $4\pi$ solid angle and has a detection efficiency
of $\gtrsim 99\%$ for both charged hadrons and photons from neutral meson
decays, fulfills all of these criteria.

A GEANT4-based simulation~\cite{geant4} was used to determine the
model-dependent corrections
required to account for the fraction of the total photoabsorption
cross section where all of the final-state particles are emitted outside the
detector acceptance. This simulation accurately
modeled the detector's geometry, composition, and electronic thresholds.

For this evaluation, differential cross sections for
the $\gamma p \rightarrow N\pi$ channels
were taken from the SAID-SM22~\cite{SM22} partial-wave analysis,
while the total cross sections for the $\gamma p \rightarrow N\pi\pi$
channels were taken from the Fix-Arenh\"ovel
model~\cite{fix2pi}, under the assumption that they proceed through
the $\gamma p \rightarrow \Delta\pi$ intermediate state.
For these last channels, the helicity asymmetry $\Delta\sigma/(\sigma_{P}+\sigma_{A})$
for all these processes was also assumed to be the same
inside and outside the detector acceptance.

The simulation confirmed that the efficiency for
partial reaction channels with at least three pions in the final state is
practically 100\% across the entire measured photon
energy range. Consequently, no acceptance corrections were applied
to these channels.

\begin{figure*}%
  \centering
  \includegraphics[scale=0.85]{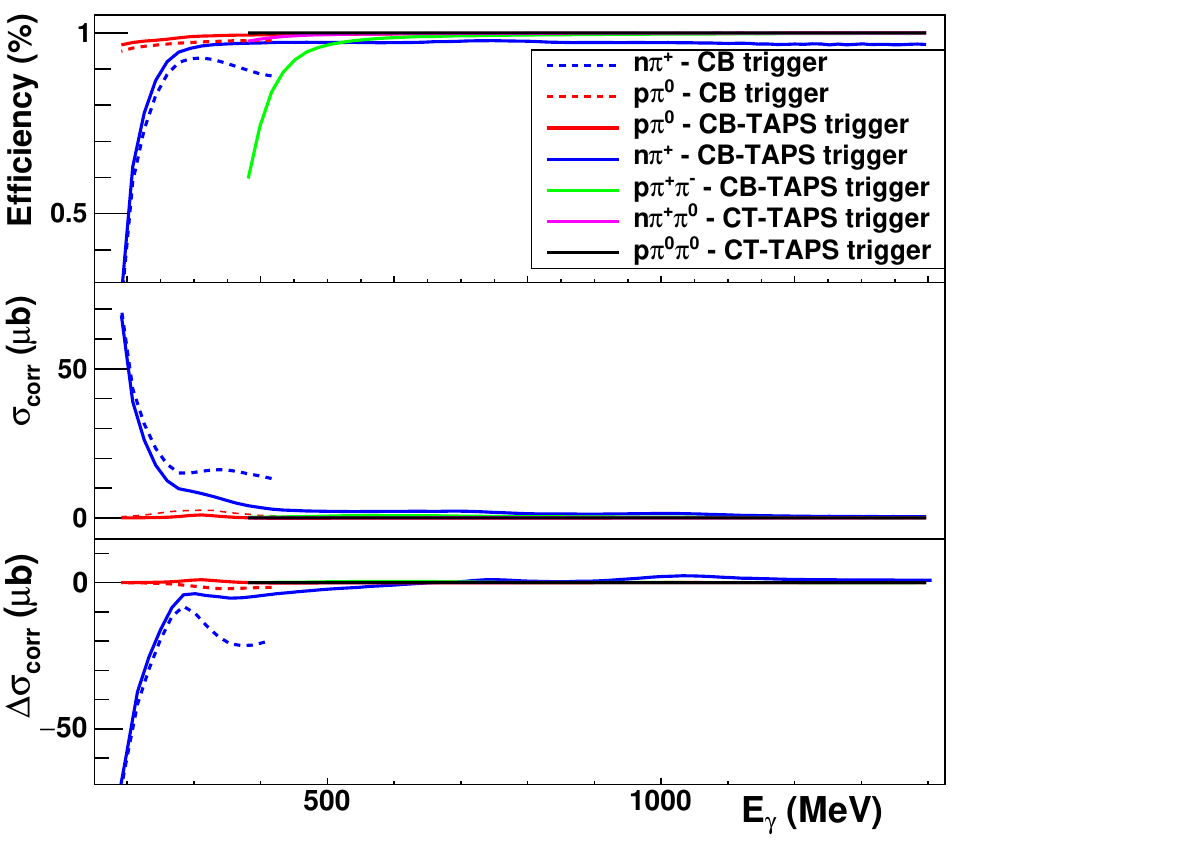}
  \caption{
    Upper plot: simulated reconstruction efficiency for
all $\gamma p \rightarrow N\pi$ and $\gamma p \rightarrow N\pi\pi$ channels
as a function of the incident photon energy.
Middle plot: the evaluated corrections to the unpolarized 
cross section $\sigma_{\text{unp}}$ are shown separately for each of
the previous channels.
Lower plot: as before but for the $\Delta\sigma$ cross section.
In all plots:
red   line: $\gamma p \rightarrow p\pi^0$;
blue  line: $\gamma p \rightarrow n\pi^+$;
green line: $\gamma p \rightarrow p\pi^+\pi^-$;
magenta line: $\gamma p \rightarrow n\pi^+\pi^0$;
black line: $\gamma p \rightarrow p\pi^0\pi^0$.
The different line styles represent the efficiency and correction values for
the CB (solid lines) and CB-TAPS (dashed lines) trigger settings,
using a 40 MeV threshold.
Above approximately 400 MeV, most of the lines become difficult to
distinguish because they converge to values of either one or zero,
depending on the plot.
}  
  \label{fig_geanteff}
\end{figure*}

The simulated reconstruction efficiency for
all $\gamma p \rightarrow N\pi$ and $\gamma p \rightarrow N\pi\pi$ channels
contributing to the unpolarized cross section
$\sigma_{\text{unp}} = (\sigma_{P} + \sigma_{A})/2$ is shown in the upper plot
of Fig.~\ref{fig_geanteff} as a function of the incident photon energy.
In the middle (lower) plot of the same figure the evaluated corrections
$\sigma_{corr}$ ($\Delta\sigma_{corr}$) to the unpolarized (polarized)
cross section are shown for each of the previous channels. 
The different line
styles represent the values  for the CB (dashed lines) and
the CB-TAPS (solid lines) trigger settings with the 40 MeV threshold.


The overall correction is mostly due to the $n\pi^+$ channel and
rapidly decreases (in absolute values)
from $E_\gamma = 200$ MeV to about $E_\gamma = 330$ MeV and remains
very low throughout the rest of the measured energy interval.

In the deuteron case, total cross sections for the $\gamma d \to pn$
and $\gamma d \to d\pi^0$ reactions were taken from the
theoretical approach of Arenh\"ovel, Fix, Schwamb (AFS)~\cite{afs},
assuming a pure phase-space distribution for both. Furthermore, for
all $\gamma d \rightarrow NN\pi(\pi)$ processes,
the dominance of quasifree processes on single nucleons was assumed.



The systematic uncertainty associated with the efficiency correction 
was evaluated by examining the different experimental thresholds and trigger conditions
applied to both the experimental and simulated data.
Simulations were also performed using $N\pi$ distributions from the BnGa-2019 partial wave analysis~\cite{boga}
and a pure phase-space distribution for the $N\pi\pi$ channels.

For the $N\pi$ channels,
a maximal efficiency difference of about 10\% was found for the $n\pi^+$ reaction compared to the previous case.
The relative uncertainty on the extrapolated values for these channels
was then conservatively estimated to be 10\% of the applied correction.

Due to the inadequate theoretical description of the experimental data for the proton in the range $500 \lesssim \hbox{E}_\gamma \lesssim 1100$~MeV (see Sec.~\ref{sec:dsigprot}), a conservative systematic uncertainty of $\pm 50\%$ was assumed for the evaluation of the missing contribution from the $N\pi\pi$ reactions, which provide the dominant contribution to the total inclusive cross section in this energy region.

The maximum absolute correction from these channels is about 0.5~$\mu$b at $\hbox{E}_\gamma \simeq 650$~MeV, corresponding to about 15\% of the total applied correction. The linear combination of this uncertainty and that of the $N\pi$ channels yields a maximal overall systematic uncertainty of $15\%$ of the applied correction.We have then conservatively assumed  this uncertainty 
to remain constant
for $\hbox{E}_\gamma > 500$~MeV.



This inclusive method was applied to evaluate the helicity-dependent cross-section difference,
$\Delta\sigma$ for both the proton and the deuteron.
For this evaluation, unpolarized background subtraction was not required due to the vanishing
contributions from the spinless carbon and oxygen nuclei.


 
\subsection{Systematic uncertainties} \label{par:syserr}

The various sources of systematic uncertainties discussed previously are
summarized in Table~\ref{tab:syserr}.

Sources of common constant systematic uncertainties come from the photon flux normalization,
the beam and target polarization, and from the target surface density.
In contrast, the systematic uncertainty related to the extrapolation correction
depends on the extrapolated value and, therefore, on E$_\gamma$.
For the proton case,
its absolute value (in rms units)  starts from $\sim 6~\mu$b at
E$_\gamma = 200$~MeV and quadratically decreases to less than 1~$\mu$b for E$_\gamma \GtrSim$~300~MeV.
The systematic uncertainty associated with the extrapolation correction for the deuteron turned out to be about twice
that estimated for the proton.

\begin{table*}[ht]
\setlength{\tabcolsep}{12pt}
\centering{}
\caption{Relative systematic uncertainties given as total widths of
  uniformly-distributed values. \label{tab:syserr}}
\begin{tabular}{l l }
 \hhline{==}
 \textbf{Source} & \textbf{Error} \\
  \hline
  Tagging efficiency & $\pm 4\%$ \\
  Beam polarization & $\pm 3\%$ \\
  Target polarization & $\pm 10\%$ \\
  Target density & $\pm 2\%$ \\
    Extrapolation correction
    E$_\gamma \leq 500$ MeV & $\pm 10\%$ of the extrapolated value \\
        Extrapolation correction
  E$_\gamma > 500$ MeV & $ \pm 15\%$ of the extrapolated value \\
%
  \hhline{==}
\end{tabular}
\end{table*}


The overall systematic uncertainty  is calculated as
the sum in quadrature of all individual uncertainties. 

\subsection{Unpolarized total inclusive cross section}

As a cross-check to validate the overall analysis procedure,
the unpolarized cross section was evaluated for both the proton and the deuteron
and compared to previously published results.

To obtain this cross section, the substantial background from the carbon and oxygen nuclei
in the target, which constitutes $\GtrSim$~75~\% of the measured raw butanol yields,
was evaluated using data from carbon runs. The normalized butanol and carbon yields
were subtracted by applying a scaling factor determined from the ratio of the live-time corrected
tagger scalers for the respective data sets.

\begin{figure*}%
    \centering
    \includegraphics[scale=0.75]{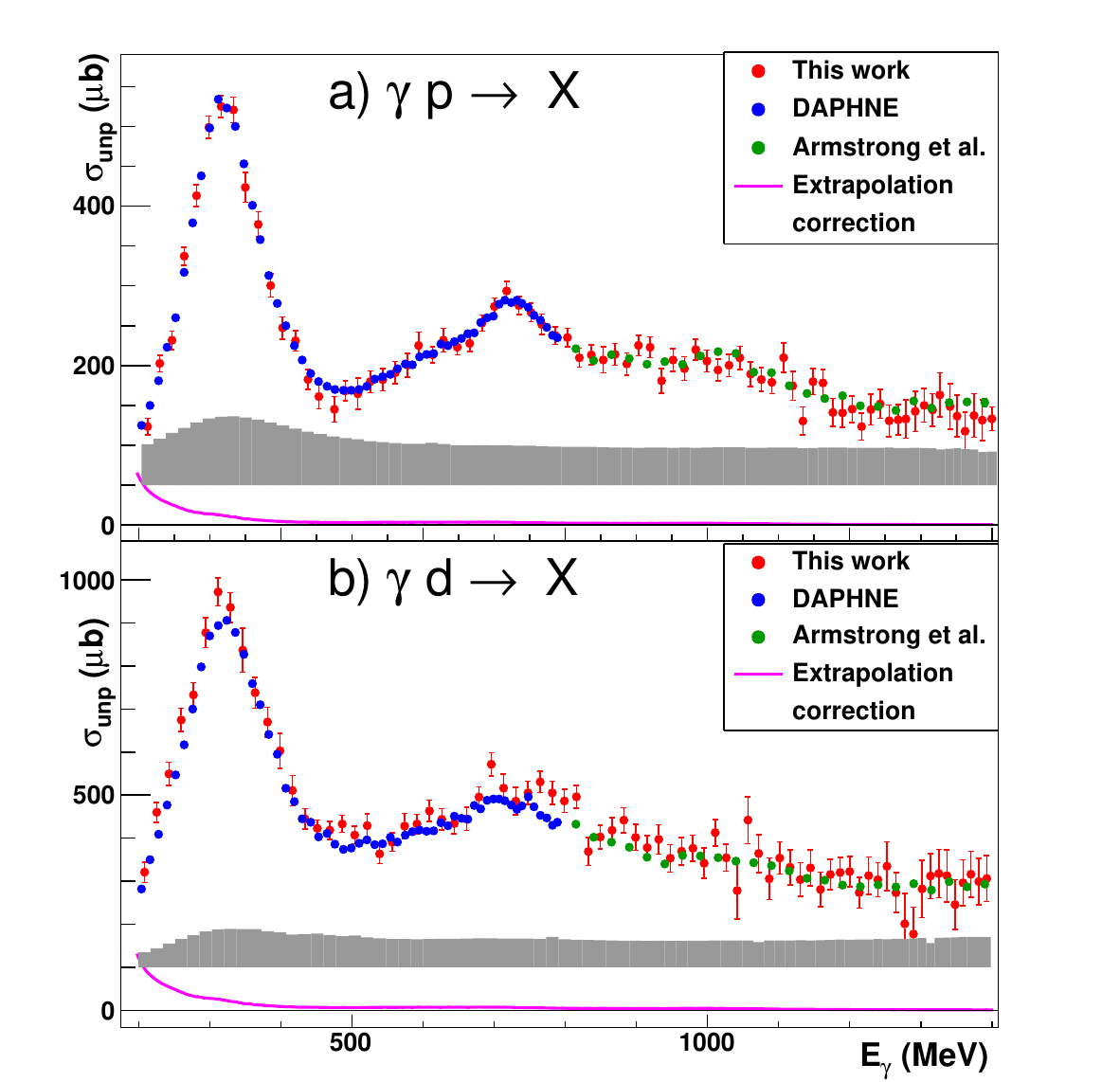}
    \caption{The unpolarized total inclusive cross section for the proton (upper plot) and deuteron (lower plot) is
      compared with previous results published by the DAPHNE
      collaboration~(\cite{maccor}), red circles, and by
      T.Armstrong {\it et al.}~(\cite{arm}), green circles.
      The solid line represents the model-dependent correction applied to account for
      detector inefficiency. The systematic uncertainties are shown as a gray bands.
    }
    \label{fig:unpoldeut}
  \end{figure*}

In Fig.~\ref{fig:unpoldeut} the values of
$\sigma_{unp}$ obtained for the deuteron are shown and compared to
previous results~\cite{maccor,arm}.

  Taking into account both the statistical and systematic uncertainties,
  the good agreement
  observed confirms the reliability of both the entire experimental
  setup and of the complete analysis procedure.


\section{Results}\label{sec:results}

\begin{figure*}%
    \centering
    \includegraphics[scale=0.75]{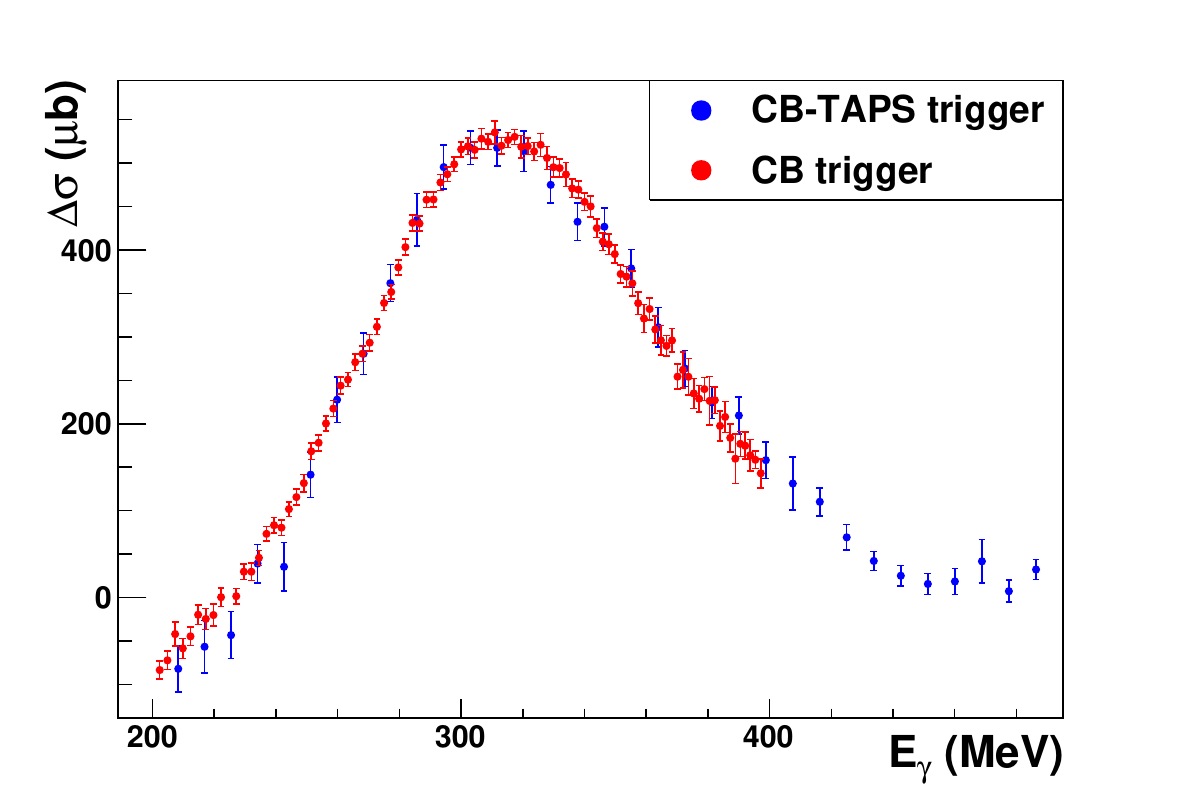}
    \caption{
      The helicity dependent total inclusive cross section $\Delta\sigma$ for the
      proton evaluated with the CB  (for E$_\gamma < 400$~MeV) and CB-TAPS trigger
      settings is shown for E$_\gamma < 500$~MeV.
    }
    \label{fig-comparison}
  \end{figure*}

Fig.~\ref{fig-comparison} shows
the helicity-dependent total inclusive cross section $\Delta\sigma$ for the proton, measured using both the CB trigger setting (for $E_\gamma < 400~\text{MeV}$) and the CB-TAPS trigger setting, in the region $E_\gamma < 500~\text{MeV}$.
%
%
The excellent agreement between these two different configurations further validates the overall analysis procedure.

For the final results, the CB trigger setting is used for photon energies below 400 MeV,
while the CB-TAPS setting is employed above this value.
Owing to the very high statistics collected, $\Delta\sigma$ for the proton
was calculated in fine photon energy bins, with a width of approximately
2 MeV for $E_\gamma < 400$ MeV and 7$-$9 MeV for $E_\gamma > 400$ MeV.

%

  \subsection{$\vec{\gamma}\vec{p}\to X$}\label{sec:dsigprot}

\begin{figure*}%
\centering
\includegraphics[scale=0.75]{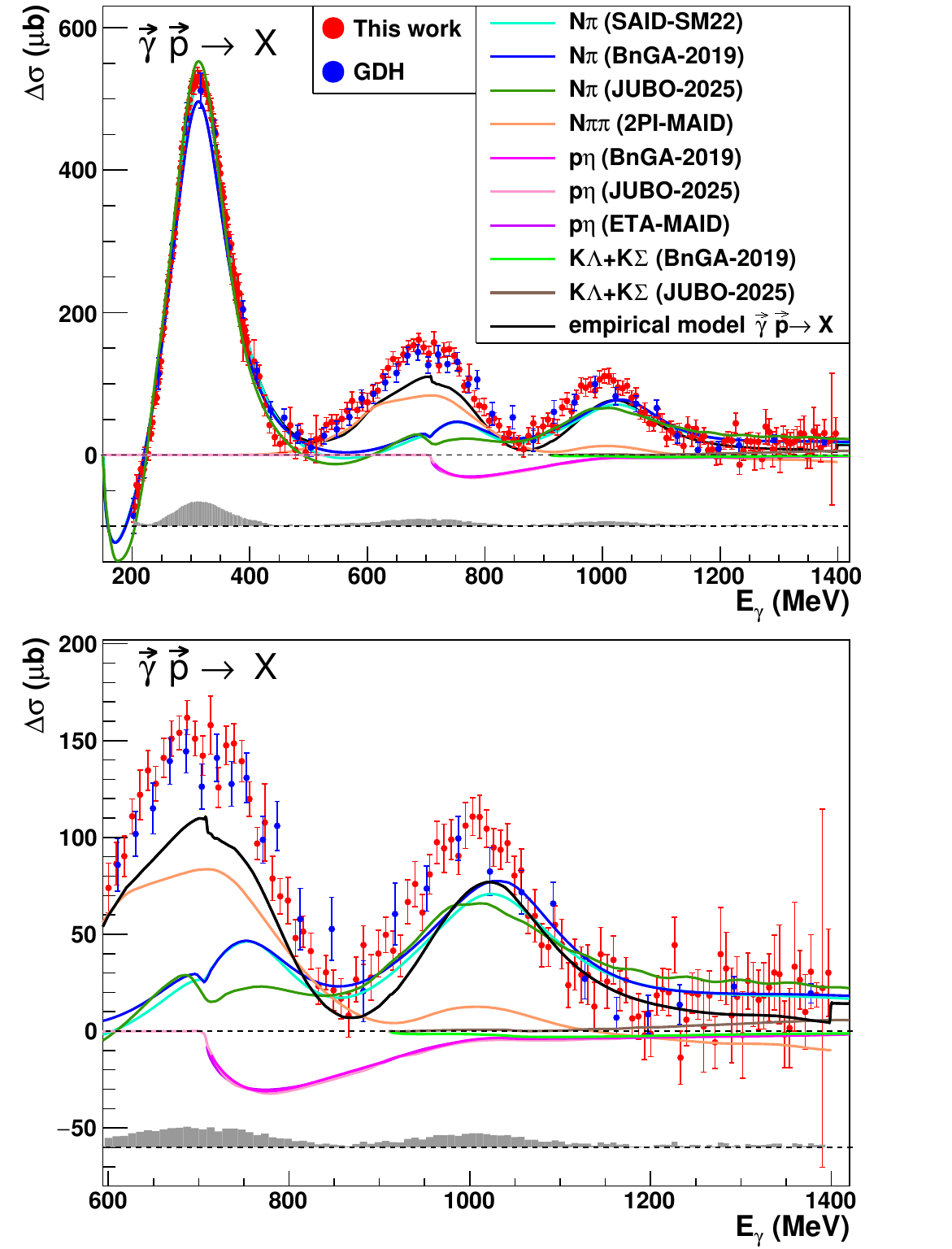}
\caption{Upper plot: the measured  inclusive helicity dependent
  cross section $\Delta\sigma$ for proton (red circles) from $E_\gamma = 200$~MeV to 1400~MeV  compared with the
  published results by the GDH collaboration~\cite{dutz04} (blue circles).
\protect\newline
  The different solid lines show the
  predictions of the relevant partial reaction channels given by different approaches:
azure line:  $\vec{\gamma} \vec{p} \to N\pi$ (SAID-SM22~\cite{SM22});
blue line: $\vec{\gamma} \vec{p} \to N\pi$ (BnGA-2019~\cite{bogak1,bogak2});
dark green line: $\vec{\gamma} \vec{p} \to N\pi$ (J\"ulich-Bonn (JuBo-2025)~\cite{jubo});
orange line:  $\vec{\gamma} \vec{p} \to N\pi\pi$ (2-PION-MAID~\cite{fix2pi}); 
magenta line:  $\vec{\gamma} \vec{p} \to N\eta$ (BnGA-2019);
pink line:  $\vec{\gamma} \vec{p} \to N\eta$ (JuBo-2025);
violet line:  $\vec{\gamma} \vec{p} \to N\eta$ (ETA-MAID~\cite{etamaid});
green line:  $\vec{\gamma} \vec{p} \to K\Lambda +K\Sigma$ (BnGA-2019~\cite{bogak1,bogak2});
brown line: $\vec{\gamma} \vec{p} \to K\Lambda +K\Sigma$ (JuBo-2025~\cite{jubo}).
%
The black line represents an empirical $\vec{\gamma}\vec{p}\to X$
prediction
obtained by adding the $N\pi$ values given by SAID-SM22, the $N\pi\pi$ 
values from MAID-2PI, the $N\eta$ values from ETA-MAID, and the  
$K\Lambda +K\Sigma$ values from BnGA-2019.
The overall contribution of the systematic uncertainties is depicted as a gray
band. 
\protect\newline
Lower plot: as before,  but from $E_\gamma = 600$~MeV to 1400~MeV.
See text for further details.
}
\label{fig:proton-full}
\end{figure*}

  The obtained total inclusive helicity dependent cross section
  $\Delta\sigma$ for the proton
is shown in the upper plot of
Fig.~\ref{fig:proton-full} (red points)
in the region from E$_\gamma$=200~MeV  to 1400~MeV and
compared to the data previously published by the GDH collaboration~\cite{dutz04} (blue points).
All numerical values of are reported in
  Table~\ref{tabprot} of the Appendix.
The lower plot of the same figure shows the subset of the obtained data
between 600~MeV and 1400~MeV.
With respect to the previous results, this
work provides a huge improvement in  statistical precision with a much finer energy binning, especially in the
$\Delta$-resonance region.

%
In the same figure,
the different solid lines show the 
predictions for all the more relevant partial reaction channels
given by different approaches.
Azure line:  $\vec{\gamma} \vec{p} \to N\pi$ - SAID-SM22~\cite{SM22};
blue line: $\vec{\gamma} \vec{p} \to N\pi$ - BnGA-2019~\cite{bogak1,bogak2};
dark green line: $\vec{\gamma} \vec{p} \to N\pi$ - J\"ulich-Bonn (JuBo-2025)~\cite{jubo});
orange line:  $\vec{\gamma} \vec{p} \to N\pi\pi$ - 2-PION-MAID~\cite{fix2pi} ;
magenta line:  $\vec{\gamma} \vec{p} \to N\eta$ - BnGA-2019;
pink line:  $\vec{\gamma} \vec{p} \to N\eta$ - JuBo-2025;
violet line:  $\vec{\gamma} \vec{p} \to N\eta$ - ETA-MAID~\cite{etamaid};
green line:  $\vec{\gamma} \vec{p} \to K\Lambda +K\Sigma$ -BnGA-2019~\cite{bogak1,bogak2};
brown line: $\vec{\gamma} \vec{p} \to K\Lambda +K\Sigma$ - JuBo-2025~\cite{jubo}.

  Below E$_\gamma \sim 500$~MeV, the predictions of the SAID-SM22 model are nearly indistinguishable from the experimental points, as they overlap almost completely. 
The  predictions for the $K\Lambda +K\Sigma$ final states are
also nearly indistinguishable in this plot, as both lie
very close to the zero y-axis value.
Likewise,
the various predictions for the $\gamma \vec{p} \to N\eta$ channel
are very similar and almost 
 completely overlap.

The black line represents an empirical $\vec{\gamma}\vec{p}\to X$
overall prediction
obtained by simply adding the $N\pi$ values given by SAID-SM22, the $N\pi\pi$ 
values from 2-PION-MAID, the $N\eta$ values from ETA-MAID, and the  
$K\Lambda +K\Sigma$ values from BnGA-2019.


The  $N\pi\pi$ prediction shown in this plot was evaluated with an updated
version of the
2PION-MAID framework presented in Ref.~\cite{fix2pi}. In this new version, 
resonance parameters (masses, hadronic and radiative widths) were not fixed
(as in the previous version) to the central values given by the
Particle Data Group (PDG) compilation. Instead,
they were treated as free parameters and 
adjusted within narrow intervals around the PDG values~\cite{ref:PDG}
in order to achieve better agreement with the existing data.

A further modification of the $\gamma N \to \pi\pi N$ amplitudes involved
the introduction of additional phenomenological (and smoothly-varying)
background terms in the $\pi\Delta$ channel for partial waves with
spin–parity $J^\pi = 3/2^\pm$ and isospin $I = 1/2$. These terms were
specifically designed to resolve the significant discrepancies previously
observed with respect to the experimental $\pi^0\pi^0$ data at
energies below the $N(1520)$ resonance peak, and were evaluated 
using the data from Refs.~\cite{Guenther2018,Ishikawa2019,Jude2022,Ghosal2023}.
Further details can be found in  Ref~\cite{Fix2024}).

%
  

In the $\Delta$(1232)-resonance region the high quality of our data allows
the small discrepancies
existing among the three different $N\pi$ approaches shown in the
previous figure to be resolved,
with the SAID-SM22 prediction reproducing our data better.
Above E$_\gamma =$500~MeV a significant discrepancy emerges between our data and the empirical prediction
obtained by summing the contribution of all the main
partial photoreaction channels.
In the so-called third 
resonance region (900~MeV~$\LessApprox \text{E}_\gamma \LessApprox$~1250~MeV),
our data peaks at a lower energy than the empirical model.

{
As mentioned before, 
this difference is mainly due to an inadequate description of the $N\pi\pi$ channels,
which provide the major contribution in this part of the measured
photon energy range.
}

A comprehensive, coupled multichannel approach is therefore required to accurately evaluate
the contributions from all relevant electromagnetic multipoles using the available
experimental database, and to improve our understanding of the properties of the higher baryon resonances.

  \subsection{$\vec{\gamma}\vec{d}\to X$}

\begin{figure*}%
\centering
\includegraphics[scale=0.75]{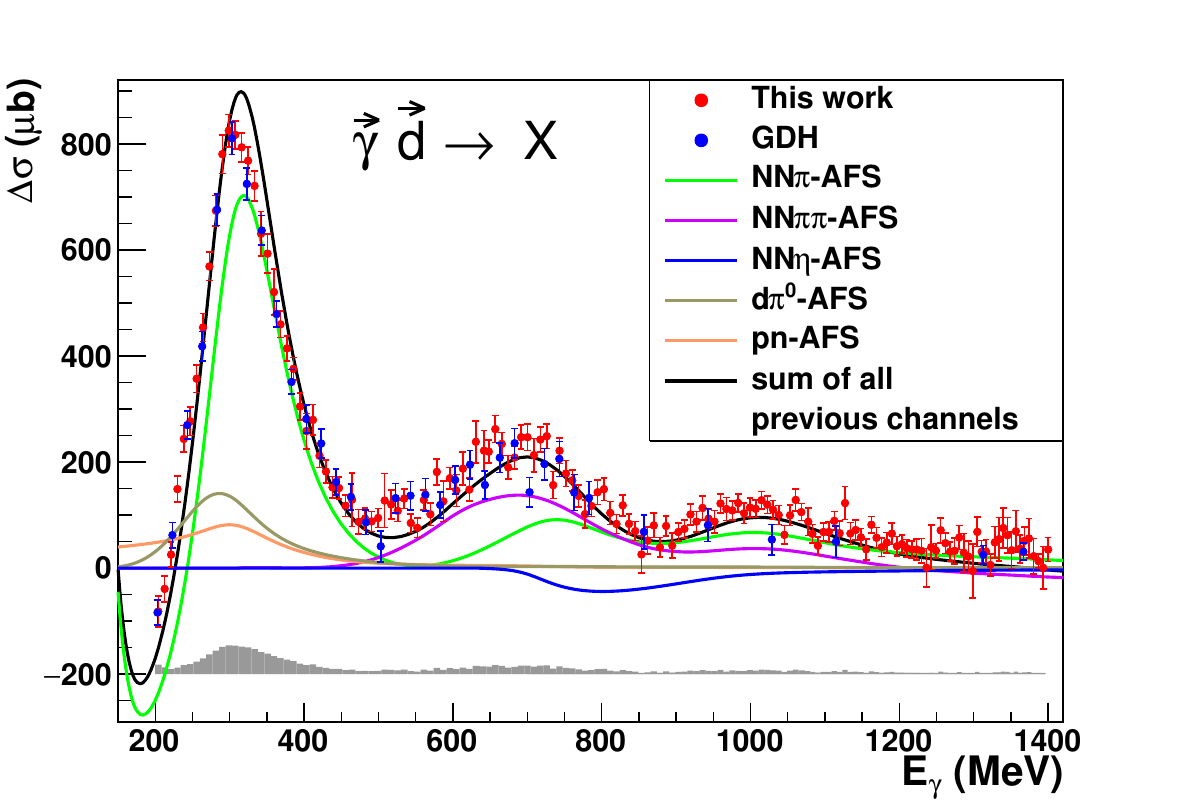}
\caption{
The new total inclusive helicity dependent
cross section $\Delta\sigma$ for the deuteron (red circles) compared
with the published results  by the GDH collaboration~(\cite{dutz05,ahr06b,ahr09}) (blue circles).
The different solid lines show the
predictions of the main partial reaction channels
given by the AFS theoretical analysis~\cite{afs}.
  Green line:   $\vec{\gamma} \vec{d} \to NN\pi$;
  violet line:  $\vec{\gamma} \vec{d} \to NN\pi\pi$;
  blue line:  $\vec{\gamma} \vec{d} \to NN\eta$;
  brownish line:  $\vec{\gamma} \vec{d} \to d\pi^0$;
  orange line: $\vec{\gamma} \vec{d} \to pn$;
  black line: sum of all previous contributions.
  The overall contribution of the systematic uncertainties is depicted as a gray
  band.
}
\label{fig:deuteron-full}
\end{figure*}

  The  total inclusive helicity dependent cross section
  $\Delta\sigma$ for the deuteron
is shown in 
Fig.~\ref{fig:deuteron-full} (red circles)
in the region from E$_\gamma$=200~MeV up to 1400~MeV and
compared to the results 
published by the GDH collaboration~\cite{dutz05,ahr06b,ahr09}
(blue circles).
All numerical values of are reported in
Table~\ref{tabneut} of the Appendix.
In this case, the present work
provides a significant improvement in  statistical precision with respect to the previous data,
with a finer energy binning especially for E$_\gamma >800$~MeV.

The different solid lines in the previous figure show the 
predictions for the main partial reaction channels
given by the theoretical AFS approach~\cite{afs}:
  green line: $\vec{\gamma} \vec{d} \to NN\pi$;
  violet line:  $\vec{\gamma} \vec{d} \to NN\pi\pi$; 
  blue line:  $\vec{\gamma} \vec{d} \to NN\eta$; 
  brownish line:  $\vec{\gamma} \vec{d} \to d\pi^0$;
  orange line:  $\vec{\gamma} \vec{d} \to pn$; 
  black line:  sum of all previous contributions.

A limitation of the calculations, which may introduce some uncertainty,
stems from the use of different theoretical frameworks for different channels.
In particular, the coherent process $\gamma d\to \pi^0 d$ is treated within
a coupled-channel approach for the $\Delta N-NN-\pi NN$ system,
whereas $\gamma d\to\pi NN$ and $\gamma d\to \pi\pi NN$ are calculated within
the impulse approximation with final-state rescattering included
perturbatively. Furthermore, for single $\pi$ and $\eta$ meson production,
the elementary operator from MAID is employed, whereas
the $\pi\pi$ photoproduction operator is derived using effective Lagrangians.

This lack of a uniform approach may be one of the reasons for the
discrepancies with the data. A more rigorous methodology is therefore
required, based on a unified and consistent theoretical framework capable
of describing all relevant channels.
  



Even considering these limitations, it is nevertheless interesting to note that the present data
confirm the apparent shift, noted in~\cite{Bass1}, in the $\Delta(1232)$ excitation contribution
to $\Delta\sigma$ {deuteron} compared to the AFS model predictions,
which employ free-space masses for the nucleon and the $\Delta$.
As shown in~\cite{Bass1}, this downwards excitation shift, $\approx -20$~MeV,
is observed in the spin-parallel but not in
the spin-antiparallel or spin averaged cross sections and
cannot be explained merely by accounting for the Fermi motion of the
target nucleons. 
%
%
This evidence points to 
the modification of the $\Delta$ excitation in nuclei.

The properties and the role of the $\Delta$ resonance 
in nuclear matter are still open puzzles
requiring further theoretical and experimental investigations.
Previous early studies suggested an attractive $\Delta$ potential
in nuclei~\cite{Ericson:1988gk}.
Chiral model studies of pion-nucleus elastic scattering gave
a $\Delta$ mass shift in medium of $-33 \times \rho/\rho_0$ MeV, where
$\rho$ is the nuclear density and $\rho_0$ is the density of nuclear matter.
Similar values, $-23 \pm 5 \times \rho/\rho_0$ MeV, were found from studies of
the $\Delta$ self-energy in nuclei using the same
theoretical framework~\cite{Oset:1987re, GarciaRecio:1989xa}.

Different interpretations have been given of coherent
$\pi^0$ photoproduction from carbon nuclei (see~\cite{Krusche:2002iq}
for  measurements performed by the TAPS collaboration at the A2-MAMI tagged photon facility).
These data have been interpreted in \cite{Peters:1998mb} as evidence for a
$\Delta$ mass reduction of 30 MeV. On the other hand, \cite{Drechsel:1999vh}
obtained a positive $\Delta$ mass shift of +20 MeV using the same data,
which they attributed to the Landau$-$Migdal effect.

In the different experimental conditions of heavy-ion collisions,
data from the SIS18 collaboration at GSI \cite{Metag:1992jh} showed that
about 30\% of the ``nucleons" behave as excited $\Delta$ resonances
(so-called resonance matter) at 2–3 times nuclear matter density
(the typical density of neutron star interiors), 
suggesting an important role for the $\Delta$ at these high densities.
 Detailed studies of
the $\Delta$ in nuclear matter will be an important part of
future experimental investigations at the GSI/FAIR facility.

%

 \subsection{$\vec{\gamma}\vec{n}\to X$}\label{Sec:peff}
  Extracting information on the neutron GDH integrand requires some
  theoretical input. Free neutron targets are not available in nature,
  necessitating measurements on neutrons bound in nuclei. Theory and
  phenomenology are then used to extract a ``free'' neutron cross section.

  In the most naive model, the rather loosely bound deuteron can be thought
  of as a combination of one free neutron and one free proton, allowing us
  in principle to extract information on the neutron GDH integrand after
  subtracting the contribution from the proton. In reality, the situation
  is complicated by the presence of various nuclear effects (Fermi motion,
  off-shell corrections, etc.). In addition, the measured cross-section
  differences also contain contributions not present for single free nucleons,
  such as coherent \(\pi^0\) production and the two-body breakup of the
  deuteron.

  Here we use the Plane Wave Impulse Approximation (PWIA) to extract the
  helicity-dependent cross section for the neutron from the deuteron and
  proton data for the higher-energy region of the measured photon energy
  interval (\(E_\gamma \gtrsim 500\)~MeV). In this region,
  as demonstrated for the \(\vec{\gamma} \vec{d} \to \pi^0 X\) channel
  in Ref.~\cite{cividini}, nuclear effects can be neglected to first approximation.
  This allows the photoproduction reaction on the deuteron to be treated
  as an incoherent sum of reactions on quasi-free protons and neutrons.

  Within this framework, the primary correction to be applied accounts for the effective nucleon
  polarization $P_E$ arising from the deuteron $D$-state contribution. This correction can be evaluated
  as (see, for instance,~\cite{rond}) 
\[
P_{E} = 1 - 1.5 P_D,
\]
where $P_D = 0.05$ is the $D$-state probability of the deuteron.

\begin{figure*}%
\centering
\includegraphics[scale=0.75]{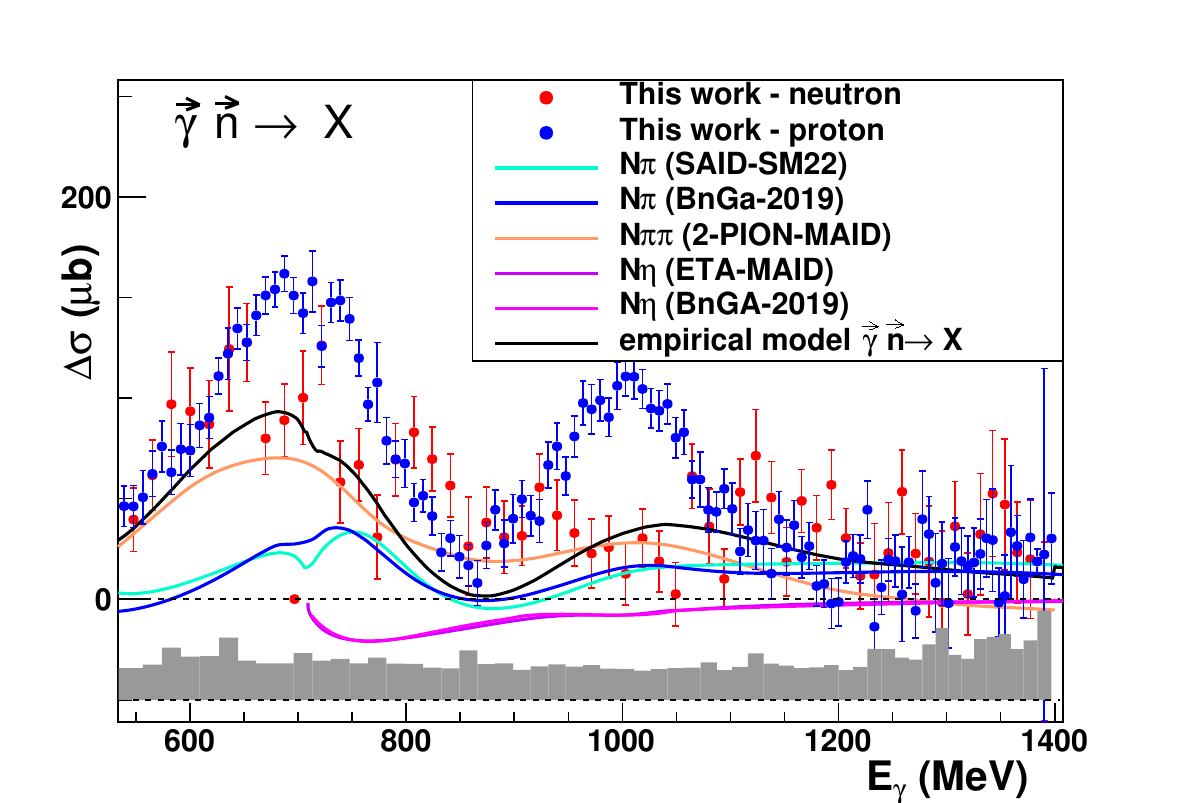}
\caption{The total inclusive helicity-dependent
  cross section $\Delta\sigma$ for the neutron (red circles),
  evaluated using Eq.~\ref{eq:neut1},
  compared with $\Delta\sigma$ for the proton (blue circles) and with
 the predictions for the relevant $\vec{\gamma}\vec{n}\to X$
 partial reaction channels given by different approaches.
 Azure line: $\vec{\gamma} \vec{n} \to N\pi$ (SAID-SM22~\cite{SM22});
 blue line: $\vec{\gamma} \vec{n} \to N\pi$ (BnGA-2019~\cite{bogak1,bogak2});
 orange line:  $\vec{\gamma} \vec{n} \to N\pi\pi$ (2-PION-MAID~\cite{fix2pi})
 (with the modifications described later);
 violet line:  $\vec{\gamma} \vec{n} \to N\eta$ (ETA-MAID~\cite{fix2pi});
 magenta line:  $\vec{\gamma} \vec{n} \to N\eta$ (BnGA-2019).
 The black line represents an empirical $\vec{\gamma}\vec{n}\to X$
overall prediction
obtained by simply adding the $N\pi$ values given by SAID-SM22, the $N\pi\pi$ 
values from MAID-2PI, and the $N\eta$ values from ETA-MAID.
  The overall contribution of the systematic uncertainties is depicted as a gray
  band. See text for further details.
}
\label{fig:neutron}
\end{figure*}

The total inclusive helicity-dependent cross section for the ``free'' neutron,
$\Delta\sigma_{\mathrm{neut}}$, is thus obtained as
\begin{equation} \label{eq:neut1}
\Delta\sigma_{\mathrm{neut}} = \Delta\sigma_{\mathrm{deut}}/P_E - \Delta\sigma_{\mathrm{prot}} .
\end{equation}
This is shown in Fig.~\ref{fig:neutron} and compared to predictions for the most relevant partial reaction
channels calculated using different approaches:
azure line: $\vec{\gamma} \vec{n} \to N\pi$ - SAID-SM22~\cite{SM22};
blue line: $\gamma p \to N\pi$ - BnGA-2019~\cite{bogak1,bogak2};
 orange line:  $\vec{\gamma} \vec{n} \to N\pi\pi$ - 2-PION-MAID~\cite{fix2pi})
 (with the modifications described earlier);
 violet line:  $\vec{\gamma} \vec{n} \to n\eta$ - ETA-MAID~\cite{fix2pi};
 magenta line:  $\gamma \vec{n} \to n\eta$ - BnGa-2019.

The predictions for the $n \eta$ final state given by the ETA-MAID and BnGa-2019 analyses are very similar and overlap almost completely in this plot.
The black line represents an empirical $\vec{\gamma}\vec{n}\to X$
overall prediction
obtained by simply adding the $N\pi$ values given by SAID-SM22, the $N\pi\pi$ 
values from MAID-2PI, and the $N\eta$ values from ETA-MAID.

The overall contribution of the systematic uncertainties is calculated as
the sum in quadrature of the deuteron and proton uncertainties, weighted
according to  Eq.~\ref{eq:neut1}.

Even though the error bars are larger than those for the proton, some
important conclusions can still be drawn from these data.
In both the so-called second 
(500~MeV~$\lesssim \text{E}_\gamma \lesssim$~900~MeV) and
third 
(900~MeV~$\lesssim \text{E}_\gamma \lesssim$~1250~MeV)
resonance regions,
$\Delta\sigma_{neutron}$ is smaller than $\Delta\sigma_{proton}$,
with the difference being more significant in the latter region.
%
This behavior
is expected, as the F$_{15}$(1680) resonance, which dominates this region,
couples more strongly to the proton than to the neutron
(see, for instance,~\cite{PDG24}).

%
In this case, the empirical $\vec{\gamma} \vec{n} \to X$ model,
represented by the black line in Fig.~\ref{fig:neutron},
seems to describe $\Delta\sigma$ more accurately for the neutron than
for the proton.
However, 
improved theory and more precise data would be needed to
go further and to draw more definitive
conclusions.

\section{Evaluation of the GDH integral}


Due to the wide energy range covered by the present, precise experimental data, a reasonable
estimate of the GDH integral (left-hand side of Eq.~\ref{eq1}) for the proton and deuteron
can be obtained by using existing models to
evaluate the missing contributions from the unmeasured energy regions.
From these estimates, the GDH integral can also be calculated for the neutron
using appropriate approximations.

\subsection{Proton}\label{sec:gdhprot}

\begin{figure*}%
\centering
\includegraphics[scale=0.75]{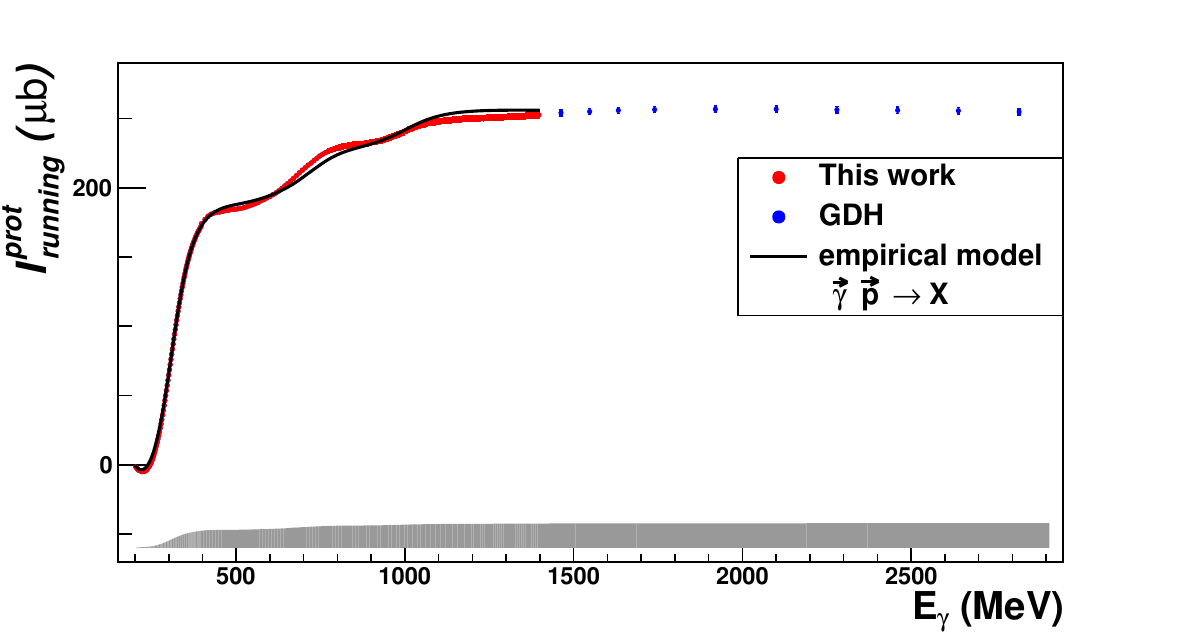}
\caption{The running GDH integral for the proton $I_{running}^{prot}$ obtained with
 the present data (red circles) and with the data from~\cite{dutz04}
  (blue circles)
   compared to the predictions of the empirical $\vec{\gamma} \vec{p} \to X$ model
  defined in Sec.~\ref{sec:dsigprot} (black solid line). 
  The gray band shows the systematic uncertainties, obtained by linearly
  adding the errors of each individual bin.
}
\label{fig:running-proton}
\end{figure*}

The diagram of Fig.~\ref{fig:running-proton}
shows the so-called “running” GDH integral:

\begin{equation}\label{eq:running}
  I_{running}= \int_{\nu_{0,exp}}^{E_\gamma}
  \frac{\Delta\sigma}{\hbox{E}_\gamma} {\hbox{dE}_\gamma}  \ ,
\end{equation}
where the infinite integral on the left hand side of Eq.~\ref{eq1}
is replaced by an integration starting
 from $\nu_{0,exp}=200$~MeV, the lowest measured
photon energy value,
to an upper limit that constitutes
the running variable.

This integral has been
obtained from the present data, ranging from 200~MeV to 1.4~GeV in photon energy, 
in combination with the higher energy data up to 2.9~GeV of
Ref.~\cite{dutz04}.
The measured values of $I_{GDH}^{prot}$ between 200 and 2900~MeV then amounts to
$255 \pm 2 \hbox{(stat)} \pm 18 \hbox{(sys)} \mu$b.

On the low-energy side (E$_\gamma < 200$~MeV), the first contribution to
photoabsorption arises from the pion production threshold ($m_{\pi}$), which occurs at about
145 MeV.
In the photon energy range between 145 and 200~MeV, the only open channels are
the single-pion production channels 
$p\pi^0$ and $n\pi^+$.
The dominant contribution in this region comes from the latter reaction, 
$n\pi^+$, driven primarily by the non-resonant
E$_0^+$ multipole.
The integration of all amplitudes across this energy range yields
a contribution of 
$-30\pm 2~\mu$b, calculated as the mean of the values obtained from 
the SAID-SM22~\cite{SM22} and BnGa-2019~\cite{boga} partial wave analyses.
The quoted systematic uncertainty corresponds to the difference
between these two estimates.

For energies above 2.9 GeV, a Regge parameterization indicates
that the contribution to the GDH integral becomes negative.
In Ref.~\cite{Bass2}, this high-energy term was
estimated to be $-15 \pm 2~\mu$b based on a fit of experimental
data for the proton's spin structure function at low $Q^2$.

Presented in the third column of Table~\ref{tab:gdhint},
the combination of experimental and model-dependent contributions produces a
result that, within the experimental uncertainties, is consistent with the GDH
sum rule value of 205~$\mu$b. This finding confirms the results of the GDH
collaboration~\cite{dutz04}.



%


\begin{table*}[ht]
\setlength{\tabcolsep}{12pt}
\caption{Estimated values of the GDH integral for proton, deuteron and
  neutron and GDH sum rule predictions for the unmeasured energy intervals
  (units: $\mu$b).
  I$_{\rm{GDH}}^{neutron}$ has been evaluated from deuteron and proton data using
  the approach outlined
  in Sec.{\ref{sec:gdhneut}}.
  The quoted errors of both the low and high-energy contributions represent the
  estimated systematic uncertainties. 
  See the article text for further explanations.
  \label{tab:gdhint}}
%
  \begin{tabular}{|c|c|c|c||c|}
    \hline
%
    &  &I$_{\rm{GDH}}^{proton}$  &I$_{\rm{GDH}}^{deuteron}$ &I$_{\rm{GDH}}^{neutron}$ \\
  \hline
    \hline
    low-energy contribution & $E_\gamma < 0.2$~GeV &$-30\pm 2$ &$-502\pm 25 $  & $-35\pm 5$\\  \hline

    %
measured & $0.2 \leq  E_\gamma \leq 2.9$~GeV& $255 \pm 2 \pm 18$  & & \\ 
interval & $0.2 \leq  E_\gamma \leq 1.8$~GeV & & $459 \pm 5 \pm 31$ & $ 234 \pm 6 \pm 37$ \\\hline    
    
%
high-energy  & $E_\gamma > 2.9 $ GeV  & $-15\pm 2$  &   & \\
contribution & $E_\gamma > 1.8 $ GeV  &   & 8 $\pm 11 $ & {$23\pm 10$}\\\hline
%
%
Total & & $ 210 \pm 2 \pm 18$ & $ -35 \pm 5 \pm 41 $ &  $222 \pm 6 \pm 38$ 
\\ \hline\hline
GDH sum rule value&  & $205$ &  $0.65$ & ${232} $\\
%
   \hline\hline
%
  \end{tabular} 
\end{table*}

\subsection{Deuteron}\label{sec:gdhdeut}

The GDH sum rule on the deuteron predicts a very small value (0.65~$\mu$b)
for the integral of Eq.~\ref{eq1} because the anomalous magnetic moment
of the deuteron is much smaller than that of the proton or
neutron (0.14~$\mu_N$ versus 1.8 and –1.9~$\mu_N$, respectively).

From the experimental point of view,
this result arises from a strong cancellation between the two-body deuteron breakup channel 
($\gamma d \to pn$), which has a very large cross-section at photon energies of a few MeV,
and the incoherent sum of inelastic channels on the individual nucleons
(e.g., $\gamma N \to N\pi(\pi)$ production at 
photon energies above $\sim$145 MeV).

\begin{figure*}%
\centering
\includegraphics[scale=0.75]{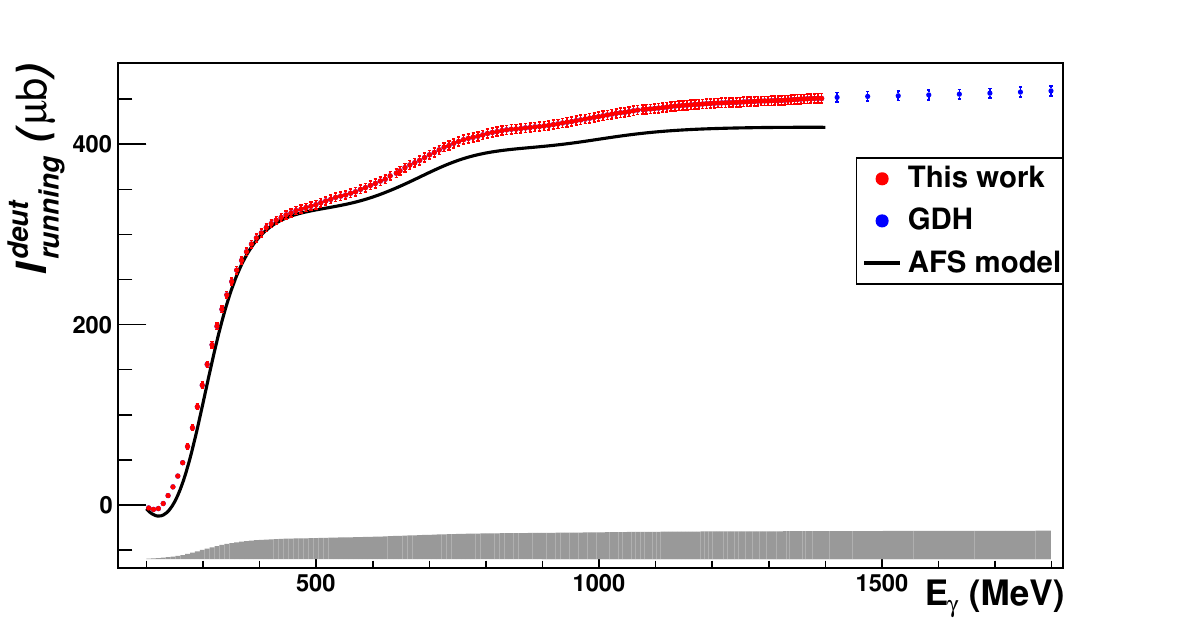}
\caption{The running GDH integral for the deuteron $I_{running}^{deut}$ obtained with
  with the present data (red circles) and with the data from~\cite{dutz05}
  (blue circles)
  is compared to
  the predictions of the AFS model~\cite{afs} (black solid line).
  The gray band shows the systematic uncertainties, obtained by linearly adding the errors of each individual bin.  
}
\label{fig:running-deuteron}
\end{figure*}

In Fig.~\ref{fig:running-deuteron},
the experimental running GDH integral for the deuteron, obtained by combining the present data (red circles)
 with the higher energy data (up to 1.8~GeV) from Ref.~\cite{dutz05} (blue circles), is 
compared to the predictions of the AFS model~(\cite{afs}). 
The experimental running integral between 200 and 1400~MeV amounts to
$459 \pm5 \pm 32~\mu$b.

The missing contribution to the GDH integral below 200 MeV has been evaluated using the
AFS model~\cite{afs}, which predicts a value of $-502$~$\mu$b.
This very large contribution is mostly concentrated in the energy region below 
E$_\gamma = 10$~MeV, where this model predicts a value of -$634$~$\mu$b.
This behavior arises because, in this energy range, the
$\gamma d\to pn$ transition
from the bound  $^3S_1$  state
 to the continuum  $^1S_0$ state 
can occur only for antiparallel photon and deuteron spins, due to the dominance of the 
$M1$ electromagnetic transition.

An indirect experimental determination of the GDH integrand for the deuteron between
the photodisintegration threshold and 6 MeV yielded a value of
$-603 \pm 43$~$\mu$b~\cite{ahmed}.
This result is in good agreement with the theoretical AFS value of
 $-627$~$\mu$b for the GDH sum rule integrated
over the same energy range.
From the comparison between the experimental and theoretical results, we estimate a
relative systematic uncertainty 
 of $\pm 5\%$ of the AFS prediction below 200~MeV.

 The contribution above  1.8~GeV can be evaluated from Regge-theory-inspired
 analyses of low-$Q^2$ photoabsorption data
 by combining  predictions from the proton and neutron
 according to Eq.~\ref{eq:neut1}.
 For the neutron,
our evaluation is based on the estimate
of $25\pm 10~\mu$b for E$_\gamma > 1.7$~GeV given by~\cite{drecrev}, which
combines the  predictions given by~\cite{bia} and~\cite{simu}
(see also Ref.~\cite{Bass:1997fh}). This value has been 
corrected to account for the experimental deuteron GDH integral value between
$\text{1.7} \leq \text{E}_\gamma \leq$1.8~GeV. The resulting 
contribution from the neutron is $23\pm 10$~$\mu$b.

The contribution for the proton above E$_\gamma = 1.7$~GeV
($-17 \pm 10~\mu$b) was evaluated in a similar way by summing up 
the Regge correction term used in the previous section with
 the experimental proton GDH integral value between
$\text{1.7} \leq \text{E}_\gamma \leq$2.9~GeV.
Finally,  applying Eq.~\ref{eq:neut1}, we obtained a high-energy
 contribution to $I_{GDH}^{deuteron}$ of $8\pm 11$~$\mu$b.


Summing up all previously-mentioned contributions
(see fourth column of Tab.~\ref{tab:gdhint}),
we obtain 
a total estimate of $I_{GDH}^{deuteron} = -35 \pm 5 \pm 41$~$\mu$b, which
is compatible (within the present systematic and statistical uncertainties)
with the deuteron GDH sum rule value.

\subsection{Neutron} \label{sec:gdhneut}

The challenge of extracting a ``free'' neutron cross section from the deuteron and proton data was discussed in Sec.~\ref{Sec:peff}.
%
For the virtual photon case, it has been shown (see, for instance,~\cite{ciofideut,deur}) that the simple assumptions leading to Eq.~\ref{eq:neut1} reproduce the experimental data well down to very low values of $Q^2$ ($Q^2 \simeq 0.01$~GeV$^2$), provided that the effects due to the $D$-state component of the deuteron wave function, which reduces the effective nucleon polarization, are taken into account.
%
In the present work, this approach was assumed to be valid also for the real photon case.
%
%
%
%
The value of the GDH integral for the neutron ($I_{GDH}^{neutron}$) was then calculated within the photon energy interval $0.2 < E_\gamma < 1.8$~GeV as
\[
\big[I^{neutron}_{GDH}\big]_{exp} \approx \frac{\big[I^{deuteron}_{GDH}\big]_{exp}}{P_E}
    - \big[I^{proton}_{GDH}\big]_{exp},
\]
where, as before, $P_E$ accounts for the nucleon depolarization caused by the deuteron $D$-state.
This then gives a value of $234 \pm 6 \pm 37$~$\mu$b.
By summing up the low-energy contribution, obtained, as in the proton case, from the SAID-SM22~\cite{SM22} and BnGa-2019~\cite{boga} partial wave analyses, and the high-energy contribution evaluated previously (see Sec.~\ref{sec:gdhdeut}), the final estimated $I^{neutron}_{GDH}$ value is also compatible with the GDH sum rule prediction (see the fourth column of Tab.~\ref{tab:gdhint}).
%


\section{Summary and conclusions}\label{sec:sum}
New precise data on the helicity-dependent total inclusive photoabsorption
cross section for both the proton and the deuteron have been obtained for
incident photon energies from 200~MeV up to  1.4~GeV.

Compared to previous results, this work provides a significant improvement
in statistical precision and a much finer energy binning.

Using these data, an estimate of the GDH integral was obtained by
employing existing models to evaluate the missing contributions from
the unmeasured energy regions. Based on these estimates,
the validity of the GDH sum rule has been verified for both
the proton and the deuteron.

From the combination of the results for the deuteron and the proton,
important information has been extracted for the free neutron.
In particular, a significant difference between the reactions
$\vec{\gamma}\vec{p}\to X$ and $\vec{\gamma}\vec{n}\to X$ was found
in the third resonance region, arising from the different strengths
of the intermediate excitation of the F$_{15}$(1680) resonance.
The value of the GDH integral for the neutron, calculated using appropriate
approximations, was also found to be compatible with the GDH sum rule
prediction.

These new data provide additional insight into nucleon properties and
establish a precise experimental benchmark for the theoretical models used
to study nucleons, both in their free state and when embedded in the nuclear
medium.

\section{Acknowledgments}

The authors wish to acknowledge the excellent support of the accelerator
group and operators of MAMI.
Thanks are also due to  C. Schneider for providing the J\"ulich-Bonn model
predictions used for comparison with our results.

This work was supported by the Deutsche Forschungsgemeinschaft (DFG, German Research Foundation) within the Research Unit “Photon-photon interactions in the Standard Model and beyond” (Project No.458854507 - FOR 5327), the European CommunityResearch Infrastructure Activity under the FP6 “Structuring the European Research Area” program (Hadron Physics, Contract No. RII3-CT-2004-506078), Schweizerischer Nationalfonds (Contract Nos. 200020-156983, 132799, 121781, 117601, 113511), the U.K. Science and
Technology Facilities Council (STFC 57071/1, 50727/1,
ST/Y000285/1),
the U.S. Department of Energy (Offices of Science and Nuclear Physics, Award Nos. DEFG02-99-ER41110, DE-FG02-88ER40415, DE-FG02-01-
ER41194) and National Science Foundation (Grant
Nos. PHY-1039130, PHY-1714833,PHY-2012940, PHY2310026, IIA-1358175), NSERC of Canada (Grant Nos.371543-2012, SAPPJ-2015-00023), INFN (Italy), the RF Ministry of Science and Higher Education, Science Program no.FSWW-2023-003.

\appendix*
\section{}
\setlength{\tabcolsep}{6pt} 
\begin{table*}[ht]
\centering{}
\caption{Numerical values of  the total inclusive helicity dependent cross section
  $\Delta\sigma$ for the proton as a function of the photon energy E$_\gamma$ in the target rest frame.
  The quoted  errors represent the 68\% confidence level of the statistical and systematic uncertainties of the experimental data, respectively. }
\begin{tabular}{cl||cl||cl||cl}
 \hhline{========}
 \textbf{E$_\gamma$} & \textbf{Value} & \textbf{E$_\gamma$} & \textbf{Value} & \textbf{E$_\gamma$} & \textbf{Value} & \textbf{E$_\gamma$} & \textbf{Value} \\
 \hline
202.3 & -83.5 $\pm$  10.5  $\pm$  7.9 & 332.0 & 494.3 $\pm$  10.4  $\pm$  32.4 & 565.3 & 62.3 $\pm$  11.4  $\pm$  4.1 & 1026.4 & 94.8 $\pm$  9.8  $\pm$  6.2 \\
204.8 & -72.4 $\pm$  10.6  $\pm$  7.4 & 334.0 & 486.9 $\pm$  14.0  $\pm$  31.9 & 574.0 & 75.9 $\pm$  12.8  $\pm$  5.0 & 1034.1 & 93.6 $\pm$  10.1  $\pm$  6.1 \\
207.3 & -42.2 $\pm$  13.9  $\pm$  5.7 & 336.0 & 470.9 $\pm$  10.7  $\pm$  30.9 & 582.7 & 63.1 $\pm$  11.1  $\pm$  4.1 & 1041.8 & 97.1 $\pm$  9.9  $\pm$  6.4 \\
209.8 & -58.6 $\pm$  11.6  $\pm$  5.7 & 338.0 & 469.5 $\pm$  9.7  $\pm$  30.8 & 591.5 & 74.6 $\pm$  12.9  $\pm$  4.9 & 1049.4 & 80.3 $\pm$  10.0  $\pm$  5.3 \\
212.3 & -44.7 $\pm$  10.4  $\pm$  5.1 & 340.0 & 455.4 $\pm$  10.2  $\pm$  29.9 & 600.3 & 73.9 $\pm$  12.9  $\pm$  4.8 & 1056.9 & 83.0 $\pm$  11.0  $\pm$  5.4 \\
214.8 & -19.9 $\pm$  11.0  $\pm$  3.8 & 342.0 & 450.2 $\pm$  11.9  $\pm$  29.5 & 609.0 & 86.4 $\pm$  11.0  $\pm$  5.7 & 1064.5 & 59.5 $\pm$  9.1  $\pm$  3.9 \\
217.3 & -24.7 $\pm$  12.0  $\pm$  4.0 & 344.0 & 425.2 $\pm$  10.8  $\pm$  27.9 & 617.7 & 90.3 $\pm$  10.5  $\pm$  5.9 & 1072.0 & 59.5 $\pm$  13.6  $\pm$  3.9 \\
219.8 & -20.2 $\pm$  12.7  $\pm$  3.4 & 345.9 & 409.6 $\pm$  10.2  $\pm$  26.9 & 626.5 & 110.9 $\pm$  9.0  $\pm$  7.3 & 1079.5 & 44.4 $\pm$  11.1  $\pm$  2.9 \\
222.2 & 0.3 $\pm$  10.6  $\pm$  3.2 & 347.9 & 406.6 $\pm$  11.7  $\pm$  26.7 & 635.2 & 122.0 $\pm$  12.8  $\pm$  8.0 & 1086.9 & 43.4 $\pm$  9.9  $\pm$  2.9 \\
227.1 & 1.5 $\pm$  8.8  $\pm$  2.5 & 349.8 & 395.3 $\pm$  10.3  $\pm$  25.9 & 643.9 & 134.6 $\pm$  10.2  $\pm$  8.8 & 1094.3 & 54.9 $\pm$  9.7  $\pm$  3.6 \\
229.6 & 29.7 $\pm$  8.8  $\pm$  3.2 & 351.7 & 372.4 $\pm$  10.1  $\pm$  24.4 & 652.6 & 127.7 $\pm$  8.9  $\pm$  8.4 & 1101.6 & 44.9 $\pm$  12.8  $\pm$  2.9 \\
232.1 & 29.7 $\pm$  10.1  $\pm$  3.0 & 353.6 & 369.3 $\pm$  11.6  $\pm$  24.2 & 661.3 & 141.1 $\pm$  10.0  $\pm$  9.3 & 1109.0 & 23.7 $\pm$  11.5  $\pm$  1.6 \\
234.5 & 45.6 $\pm$  8.6  $\pm$  3.7 & 355.5 & 361.6 $\pm$  14.4  $\pm$  23.7 & 670.0 & 151.0 $\pm$  9.2  $\pm$  9.9 & 1116.2 & 34.4 $\pm$  13.1  $\pm$  2.3 \\
236.9 & 73.3 $\pm$  8.4  $\pm$  5.2 & 357.4 & 338.8 $\pm$  13.2  $\pm$  22.2 & 678.7 & 154.0 $\pm$  8.8  $\pm$  10.1 & 1123.5 & 29.1 $\pm$  14.4  $\pm$  1.9 \\
239.4 & 83.2 $\pm$  9.1  $\pm$  5.8 & 359.3 & 321.1 $\pm$  16.3  $\pm$  21.1 & 687.3 & 161.8 $\pm$  8.8  $\pm$  10.6 & 1130.6 & 29.1 $\pm$  15.4  $\pm$  1.9 \\
241.8 & 80.4 $\pm$  8.9  $\pm$  5.6 & 361.1 & 332.1 $\pm$  12.6  $\pm$  21.8 & 696.0 & 151.0 $\pm$  8.9  $\pm$  9.9 & 1137.8 & 12.7 $\pm$  12.6  $\pm$  0.8 \\
244.2 & 101.7 $\pm$  8.5  $\pm$  6.9 & 363.0 & 308.5 $\pm$  15.3  $\pm$  20.2 & 704.6 & 142.2 $\pm$  10.1  $\pm$  9.3 & 1144.9 & 39.7 $\pm$  13.8  $\pm$  2.6 \\
246.6 & 115.5 $\pm$  9.1  $\pm$  7.8 & 364.8 & 296.0 $\pm$  17.1  $\pm$  19.4 & 713.2 & 158.0 $\pm$  15.1  $\pm$  10.4 & 1152.0 & 25.7 $\pm$  13.8  $\pm$  1.7 \\
249.0 & 131.6 $\pm$  9.9  $\pm$  8.8 & 366.6 & 289.6 $\pm$  11.7  $\pm$  19.0 & 721.9 & 125.8 $\pm$  10.4  $\pm$  8.2 & 1159.0 & 36.8 $\pm$  12.4  $\pm$  2.4 \\
251.4 & 168.3 $\pm$  9.4  $\pm$  11.1 & 368.4 & 295.9 $\pm$  13.5  $\pm$  19.4 & 730.5 & 147.5 $\pm$  10.0  $\pm$  9.7 & 1165.9 & 20.8 $\pm$  10.4  $\pm$  1.4 \\
253.8 & 178.1 $\pm$  9.2  $\pm$  11.8 & 370.2 & 254.2 $\pm$  14.4  $\pm$  16.7 & 739.0 & 148.5 $\pm$  10.2  $\pm$  9.7 & 1172.9 & 26.4 $\pm$  11.6  $\pm$  1.7 \\
256.2 & 200.3 $\pm$  8.5  $\pm$  13.2 & 371.9 & 261.9 $\pm$  20.6  $\pm$  17.2 & 747.6 & 139.4 $\pm$  10.7  $\pm$  9.1 & 1179.8 & 6.6 $\pm$  10.1  $\pm$  0.4 \\
258.6 & 217.4 $\pm$  9.3  $\pm$  14.3 & 373.7 & 254.1 $\pm$  20.9  $\pm$  16.7 & 756.2 & 119.8 $\pm$  8.9  $\pm$  7.9 & 1186.6 & 7.6 $\pm$  11.8  $\pm$  0.5 \\
261.0 & 243.8 $\pm$  9.7  $\pm$  16.0 & 375.4 & 234.9 $\pm$  17.5  $\pm$  15.4 & 764.8 & 96.8 $\pm$  8.3  $\pm$  6.4 & 1193.4 & -2.2 $\pm$  12.4  $\pm$  0.2 \\
263.4 & 250.8 $\pm$  8.1  $\pm$  16.5 & 377.2 & 228.8 $\pm$  15.4  $\pm$  15.0 & 773.3 & 107.7 $\pm$  19.9  $\pm$  7.1 & 1200.1 & -1.4 $\pm$  11.9  $\pm$  0.1 \\
265.7 & 270.8 $\pm$  9.4  $\pm$  17.8 & 378.9 & 239.9 $\pm$  13.2  $\pm$  15.7 & 781.8 & 78.8 $\pm$  11.4  $\pm$  5.2 & 1206.8 & 18.5 $\pm$  10.3  $\pm$  1.2 \\
268.1 & 280.7 $\pm$  9.2  $\pm$  18.4 & 380.6 & 226.3 $\pm$  27.9  $\pm$  14.8 & 790.3 & 69.6 $\pm$  9.1  $\pm$  4.6 & 1213.5 & 21.5 $\pm$  10.7  $\pm$  1.4 \\
270.4 & 293.3 $\pm$  9.3  $\pm$  19.2 & 382.3 & 227.1 $\pm$  15.1  $\pm$  14.9 & 798.8 & 67.5 $\pm$  12.0  $\pm$  4.4 & 1220.1 & 20.0 $\pm$  11.9  $\pm$  1.3 \\
272.7 & 311.6 $\pm$  8.8  $\pm$  20.5 & 383.9 & 197.5 $\pm$  17.1  $\pm$  13.0 & 807.2 & 48.1 $\pm$  9.5  $\pm$  3.2 & 1226.7 & 44.4 $\pm$  14.3  $\pm$  2.9 \\
275.0 & 339.0 $\pm$  8.7  $\pm$  22.2 & 385.6 & 207.8 $\pm$  17.8  $\pm$  13.6 & 815.7 & 51.4 $\pm$  8.0  $\pm$  3.4 & 1233.2 & -13.7 $\pm$  13.8  $\pm$  0.9 \\
277.4 & 351.7 $\pm$  8.4  $\pm$  23.1 & 387.2 & 183.8 $\pm$  16.0  $\pm$  12.1 & 824.1 & 41.3 $\pm$  9.4  $\pm$  2.7 & 1239.6 & 5.9 $\pm$  13.5  $\pm$  0.4 \\
279.7 & 380.0 $\pm$  8.5  $\pm$  24.9 & 388.9 & 159.8 $\pm$  28.6  $\pm$  10.5 & 832.5 & 23.3 $\pm$  9.3  $\pm$  1.5 & 1246.1 & 19.6 $\pm$  10.5  $\pm$  1.3 \\
282.0 & 403.3 $\pm$  9.2  $\pm$  26.5 & 390.5 & 176.9 $\pm$  14.3  $\pm$  11.6 & 840.8 & 30.3 $\pm$  8.8  $\pm$  2.0 & 1252.4 & 18.9 $\pm$  12.8  $\pm$  1.2 \\
284.3 & 431.2 $\pm$  9.3  $\pm$  28.3 & 392.1 & 174.7 $\pm$  15.6  $\pm$  11.5 & 849.2 & 21.1 $\pm$  10.5  $\pm$  1.4 & 1258.8 & 2.4 $\pm$  23.5  $\pm$  0.2 \\
286.5 & 430.4 $\pm$  8.7  $\pm$  28.2 & 393.7 & 163.8 $\pm$  18.0  $\pm$  10.7 & 857.6 & 16.8 $\pm$  10.4  $\pm$  1.1 & 1265.1 & 18.4 $\pm$  12.7  $\pm$  1.2 \\
288.8 & 457.9 $\pm$  8.9  $\pm$  30.0 & 395.4 & 158.5 $\pm$  10.2  $\pm$  10.4 & 865.9 & 8.1 $\pm$  11.2  $\pm$  0.5 & 1271.3 & -5.8 $\pm$  13.4  $\pm$  0.4 \\
291.1 & 458.2 $\pm$  8.9  $\pm$  30.0 & 397.1 & 142.9 $\pm$  16.7  $\pm$  9.4 & 874.2 & 26.7 $\pm$  11.5  $\pm$  1.7 & 1277.5 & 39.8 $\pm$  24.1  $\pm$  2.6 \\
293.3 & 477.8 $\pm$  9.3  $\pm$  31.3 & 407.5 & 131.2 $\pm$  30.5  $\pm$  8.6 & 882.4 & 44.5 $\pm$  10.0  $\pm$  2.9 & 1283.6 & 32.3 $\pm$  18.7  $\pm$  2.1 \\
295.6 & 487.2 $\pm$  8.4  $\pm$  32.0 & 416.3 & 110.1 $\pm$  16.1  $\pm$  7.2 & 890.7 & 27.7 $\pm$  11.8  $\pm$  1.8 & 1289.7 & 8.1 $\pm$  24.5  $\pm$  0.5 \\
297.8 & 498.7 $\pm$  8.5  $\pm$  32.7 & 425.0 & 69.2 $\pm$  14.6  $\pm$  4.5 & 898.9 & 40.0 $\pm$  12.0  $\pm$  2.6 & 1295.8 & 17.7 $\pm$  20.8  $\pm$  1.2 \\
300.0 & 515.8 $\pm$  8.5  $\pm$  33.8 & 433.8 & 42.0 $\pm$  11.0  $\pm$  2.8 & 907.1 & 49.7 $\pm$  9.2  $\pm$  3.3 & 1301.8 & -2.0 $\pm$  22.3  $\pm$  0.2 \\
302.2 & 519.4 $\pm$  9.3  $\pm$  34.1 & 442.5 & 25.0 $\pm$  11.6  $\pm$  1.7 & 915.2 & 41.5 $\pm$  10.0  $\pm$  2.7 & 1307.7 & 25.9 $\pm$  14.5  $\pm$  1.7 \\
304.4 & 515.3 $\pm$  9.2  $\pm$  33.8 & 451.3 & 15.6 $\pm$  12.0  $\pm$  1.0 & 923.4 & 38.8 $\pm$  10.6  $\pm$  2.5 & 1313.7 & 18.9 $\pm$  16.6  $\pm$  1.2 \\
306.6 & 528.2 $\pm$  11.7  $\pm$  34.6 & 460.1 & 18.4 $\pm$  15.0  $\pm$  1.2 & 931.5 & 66.7 $\pm$  11.3  $\pm$  4.4 & 1319.6 & 16.1 $\pm$  15.3  $\pm$  1.1 \\
308.8 & 524.5 $\pm$  8.8  $\pm$  34.4 & 468.8 & 41.6 $\pm$  25.0  $\pm$  2.7 & 939.6 & 76.0 $\pm$  11.6  $\pm$  5.0 & 1325.4 & 18.3 $\pm$  17.1  $\pm$  1.2 \\
311.0 & 530.4 $\pm$  13.3  $\pm$  35.1 & 477.6 & 7.2 $\pm$  12.6  $\pm$  0.5 & 947.6 & 61.2 $\pm$  9.4  $\pm$  4.0 & 1331.2 & 22.5 $\pm$  17.5  $\pm$  1.5 \\
313.1 & 530.0 $\pm$  9.7  $\pm$  34.1 & 486.4 & 32.2 $\pm$  11.6  $\pm$  2.1 & 955.6 & 80.9 $\pm$  10.2  $\pm$  5.3 & 1336.9 & 30.3 $\pm$  19.3  $\pm$  2.0 \\
315.2 & 530.6 $\pm$  8.7  $\pm$  34.5 & 495.1 & 3.9 $\pm$  11.8  $\pm$  0.3 & 963.6 & 97.5 $\pm$  10.9  $\pm$  6.4 & 1342.6 & 29.4 $\pm$  25.5  $\pm$  1.9 \\
317.4 & 525.3 $\pm$  8.6  $\pm$  34.8 & 503.9 & 16.4 $\pm$  13.1  $\pm$  1.1 & 971.6 & 94.4 $\pm$  12.4  $\pm$  6.2 & 1348.2 & -1.7 $\pm$  17.2  $\pm$  0.1 \\
319.5 & 518.9 $\pm$  13.1  $\pm$  34.0 & 512.7 & 22.1 $\pm$  33.1  $\pm$  1.5 & 979.5 & 98.9 $\pm$  10.3  $\pm$  6.5 & 1353.8 & 1.6 $\pm$  20.4  $\pm$  0.1 \\
321.6 & 519.7 $\pm$  9.5  $\pm$  34.1 & 521.4 & 24.1 $\pm$  11.6  $\pm$  1.6 & 987.4 & 90.4 $\pm$  9.7  $\pm$  5.9 & 1359.4 & 33.3 $\pm$  33.3  $\pm$  2.2 \\
323.7 & 513.4 $\pm$  10.2  $\pm$  33.7 & 530.2 & 42.4 $\pm$  10.1  $\pm$  2.8 & 995.3 & 106.1 $\pm$  11.9  $\pm$  7.0 & 1365.2 & 26.5 $\pm$  15.2  $\pm$  1.7 \\
325.8 & 521.1 $\pm$  13.3  $\pm$  34.2 & 539.0 & 46.3 $\pm$  10.1  $\pm$  3.0 & 1003.1 & 110.7 $\pm$  9.8  $\pm$  7.3 & 1371.2 & 9.9 $\pm$  19.3  $\pm$  0.7 \\
327.9 & 506.0 $\pm$  13.1  $\pm$  33.2 & 547.7 & 46.1 $\pm$  10.4  $\pm$  3.0 & 1010.9 & 110.6 $\pm$  11.2  $\pm$  7.3 & 1377.4 & 30.7 $\pm$  19.0  $\pm$  2.0 \\
329.9 & 495.3 $\pm$  11.5  $\pm$  32.5 & 556.5 & 50.7 $\pm$  13.2  $\pm$  3.3 & 1018.6 & 104.5 $\pm$  9.7  $\pm$  6.9 & 1383.8 & 18.7 $\pm$  6.3  $\pm$  1.2 \\
1390.2& 22.2$\pm$ 92.5$\pm$  1.5 &  1397.0 &   30.2$\pm$  22.6 $\pm$ 2.0 & &  & & \\

%
    \hhline{========}
\end{tabular} \label{tabprot}
  \end{table*}

\begin{table*}[ht]
\centering{}
\caption{Numerical values of  the total inclusive helicity dependent cross section
  $\Delta\sigma$ for the deuteron as a function of the photon energy E$_\gamma$ in the target rest frame.
  The quoted  errors represent the 68\% confidence level of the statistical and systematic uncertainties of the experimental data, respectively. }
\begin{tabular}{cl||cl||cl||cl}
 \hhline{========}
 \textbf{E$_\gamma$} & \textbf{Value} & \textbf{E$_\gamma$} & \textbf{Value} & \textbf{E$_\gamma$} & \textbf{Value} & \textbf{E$_\gamma$} & \textbf{Value} \\
 \hline
204.1 & -81.9 $\pm$  29.7  $\pm$  12.7 & 534.6 & 131.1$\pm$  18.6   $\pm$  8.6 & 861.7 & 51.9   $\pm$  25.3  $\pm$  3.4 & 1155.5 & 35.5 $\pm$  19.7  $\pm$  2.3 \\
212.6 & -39.2 $\pm$  24.6  $\pm$  8.8 & 543.4 & 85.0  $\pm$  17.5   $\pm$  5.6 & 870.0 & 80.4   $\pm$  23.1  $\pm$  5.3 & 1162.4 & 81.7 $\pm$  16.0  $\pm$  5.4 \\
221.2 & 25.3  $\pm$  23.7  $\pm$  6.5 & 552.1 & 75.7  $\pm$  18.3   $\pm$  5.0 & 878.3 & 39.1   $\pm$  18.3  $\pm$  2.6 & 1169.4 & 57.2 $\pm$  19.2  $\pm$  3.8 \\
229.7 & 149.0 $\pm$  24.4  $\pm$  11 & 560.9 & 128.6  $\pm$  18.8   $\pm$  8.4 & 886.5 & 79.3   $\pm$  18.5  $\pm$  5.2 & 1176.3 & 39.3 $\pm$  17.5  $\pm$  2.6 \\
238.3 & 243.6 $\pm$  25.2  $\pm$  16.5 & 569.6 & 101.5 $\pm$  20.9  $\pm$  6.7 & 894.8 & 42.4   $\pm$  23.5  $\pm$  2.8 & 1183.2 & 48.1 $\pm$  19.2  $\pm$  3.2 \\
246.9 & 276.8 $\pm$  26.8  $\pm$  18.5 & 578.4 & 181.3 $\pm$  24.7  $\pm$  11.9 & 903.0 & 68.0  $\pm$  16.9  $\pm$  4.5 & 1190.0 & 64.9 $\pm$  20.6  $\pm$  4.3 \\
255.5 & 357.2 $\pm$  25.8  $\pm$  23.6 & 587.1 & 125.9 $\pm$  38.0  $\pm$  8.3 & 911.2 & 73.9   $\pm$  17.0  $\pm$  4.9 & 1196.8 & 41.1 $\pm$  18.3  $\pm$  2.7 \\
264.1 & 453.8 $\pm$  26.7  $\pm$  29.9 & 595.9 & 169.5 $\pm$  20.1  $\pm$  11.1 & 919.3 & 101.2 $\pm$  19.4  $\pm$  6.6 & 1203.5 & 45.2 $\pm$  16.8  $\pm$  3.0 \\
272.7 & 568.9 $\pm$  26.8  $\pm$  37.4 & 604.6 & 146.1 $\pm$  30.2  $\pm$  9.6 & 927.4 & 87.6   $\pm$  19.3  $\pm$  5.8 & 1210.2 & 37.4 $\pm$  16.2  $\pm$  2.5 \\
281.4 & 673.9 $\pm$  28.6  $\pm$  44.3 & 613.4 & 187.5 $\pm$  31.2  $\pm$  12.3 & 935.5 & 113.4 $\pm$  22.7  $\pm$  7.4 & 1216.8 & 35.2 $\pm$  16.1  $\pm$  2.3 \\
290.1 & 780.6 $\pm$  36.4  $\pm$  51.3 & 622.1 & 148.0 $\pm$  21.5  $\pm$  9.7 & 943.6 & 93.2   $\pm$  18.3  $\pm$  6.1 & 1223.4 & 35.1 $\pm$  19.8  $\pm$  2.3 \\
298.7 & 824.9 $\pm$  30.8  $\pm$  54.2 & 630.8 & 238.1 $\pm$  39.8  $\pm$  15.6 & 951.6 & 87.4  $\pm$  20.9  $\pm$  5.7 & 1229.9 & 32.5 $\pm$  23.4  $\pm$  2.1 \\
307.4 & 817.0 $\pm$  26.3  $\pm$  53.6 & 641.7 & 221.4 $\pm$  38.4  $\pm$  14.5 & 959.6 & 121.7 $\pm$  17.7  $\pm$  8.0 & 1236.4 & 0.7  $\pm$  36.3  $\pm$  0.2 \\
316.0 & 793.4 $\pm$  27.3  $\pm$  52.1 & 648.2 & 219.5 $\pm$  21.8  $\pm$  14.4 & 967.6 & 111.4 $\pm$  21.4  $\pm$  4.7 & 1242.9 & 39.4 $\pm$  41.8  $\pm$  2.6 \\
324.7 & 768.2 $\pm$  26.2  $\pm$  50.4 & 656.9 & 262.1 $\pm$  25.7  $\pm$  17.2 & 975.5 & 108.8 $\pm$  18.2  $\pm$  7.1 & 1249.3 & 32.7 $\pm$  13.4  $\pm$  2.2 \\
333.4 & 720.7 $\pm$  28.1  $\pm$  47.3 & 665.6 & 234.2 $\pm$  21.8  $\pm$  15.4 & 983.5 & 122.7 $\pm$  16.8  $\pm$  6.1 & 1255.6 & 71.2 $\pm$  23.6  $\pm$  4.7 \\
342.1 & 630.5 $\pm$  41.7  $\pm$  41.4 & 674.3 & 189.9 $\pm$  22.2  $\pm$  12.5 & 991.3 & 103.4 $\pm$  19.0  $\pm$  5.5 & 1261.9 & 47.0 $\pm$  19.5  $\pm$  3.1 \\
350.8 & 593.0 $\pm$  36.6  $\pm$  38.9 & 683.0 & 208.5 $\pm$  21.8  $\pm$  13.7 & 999.2 & 114.2 $\pm$  18.3  $\pm$  7.5 & 1268.2 & 31.0 $\pm$  17.8  $\pm$  2.0 \\
359.5 & 520.5 $\pm$  43.6  $\pm$  34.2 & 691.6 & 246.8 $\pm$  22.6  $\pm$  16.2 & 1007.0 & 111.8$\pm$  16.7  $\pm$  6.7 & 1274.4 & 32.4 $\pm$  24.6  $\pm$  2.1 \\
368.2 & 459.6 $\pm$  24.9  $\pm$  30.2 & 700.3 & 246.8 $\pm$  25.4  $\pm$  16.2 & 1014.8 & 128  $\pm$  16.3  $\pm$  8.4 & 1280.6 & 57.6 $\pm$  22.8  $\pm$  3.8 \\
376.9 & 414.3 $\pm$  23.4  $\pm$  27.2 & 708.9 & 213.0 $\pm$  31.4  $\pm$  14 & 1022.5 & 119.7  $\pm$  16.5  $\pm$  7.9 & 1286.7 & 29.0 $\pm$  35.9  $\pm$  1.9 \\
385.7 & 375.6 $\pm$  21.2  $\pm$  24.7 & 717.6 & 242.3 $\pm$  23.3  $\pm$  15.9 & 1030.2 & 110.1$\pm$  17.2  $\pm$  7.2 & 1292.8 & 21.6 $\pm$  29.9  $\pm$  1.4 \\
394.4 & 304.6 $\pm$  25.7  $\pm$  20.0 & 726.2 & 248.8 $\pm$  22.9  $\pm$  16.3 & 1037.9 & 100.1$\pm$  18.0  $\pm$  6.6 & 1298.8 & -5.5 $\pm$  51.0  $\pm$  0.4 \\
403.2 & 258.5 $\pm$  33.5  $\pm$  17.0 & 734.8 & 156.5 $\pm$  25.6  $\pm$  10.3 & 1045.6 & 62.6 $\pm$  17.3  $\pm$  4.1 & 1304.8 & 68.3 $\pm$  27.6  $\pm$  4.5 \\
411.9 & 279.5 $\pm$  28.8  $\pm$  18.4 & 743.3 & 221.3 $\pm$  24.2  $\pm$  14.5 & 1053.1 & 99.5 $\pm$  19.3  $\pm$  6.5 & 1310.7 & 30.6 $\pm$  21.2  $\pm$  2.0 \\
420.7 & 211.8 $\pm$  22.1  $\pm$  13.9 & 751.9 & 178.3 $\pm$  24.2  $\pm$  11.7 & 1060.7 & 128.5$\pm$  20.8  $\pm$  8.4 & 1316.6 & 31.1 $\pm$  23.1  $\pm$  2.0 \\
429.4 & 182.0 $\pm$  21.3  $\pm$  12.0 & 760.5 & 165.3 $\pm$  19.6  $\pm$  10.9 & 1068.3 & 105.6$\pm$  15.8  $\pm$  6.9 & 1322.5 & 5.8  $\pm$  20.8  $\pm$  0.4 \\
438.2 & 152.2 $\pm$  20.4  $\pm$  10.0 & 769.0 & 136.1 $\pm$  20.4  $\pm$  8.9 & 1077.6 & 87.4  $\pm$  25.8  $\pm$  5.7 & 1328.3 & 48.5 $\pm$  29.5  $\pm$  3.2 \\
446.9 & 150.9 $\pm$  20.7  $\pm$  9.9  & 777.5 & 101.8 $\pm$  26.4  $\pm$  6.7 & 1083.2 & 64.7  $\pm$  17.2  $\pm$  4.3 & 1334.0 & 54.5 $\pm$  41.0  $\pm$  3.6 \\
455.7 & 117.6 $\pm$  20.1  $\pm$  7.7  & 786.0 & 117.2   $\pm$  19.8  $\pm$  7.7 & 1090.6 & 42.2  $\pm$  16.7  $\pm$  2.8 & 1339.7 & 75.5 $\pm$  37.6  $\pm$  5.0 \\
464.4 & 129.0 $\pm$  50.6  $\pm$  8.5 & 794.5 & 143.0  $\pm$  23.0  $\pm$  9.4 & 1098.0 & 68.0  $\pm$  16.6  $\pm$  4.5 & 1345.4 & 61.9 $\pm$  33.4  $\pm$  4.1 \\
473.2 & 87.4  $\pm$  23.4  $\pm$  5.7 & 803.0 & 148.9    $\pm$  21.3  $\pm$  9.8 & 1105.3 & 68.1  $\pm$  19.0  $\pm$  4.5 & 1351.0 & 33.6 $\pm$  34.2  $\pm$  2.2 \\
482.0 & 93.3  $\pm$  20.4  $\pm$  6.1 & 811.5 & 104.0  $\pm$  21.7  $\pm$  6.8 & 1112.6 & 89.0  $\pm$  17.4  $\pm$  5.9 & 1356.6 & 69.4 $\pm$  39.3  $\pm$  4.6 \\
490.8 & 87.9  $\pm$  19.2  $\pm$  5.8 & 819.9 & 83.2   $\pm$  18.3  $\pm$  5.5 & 1119.9 & 65.6  $\pm$  22.7  $\pm$  4.3 & 1362.3 & 36.3 $\pm$  26.5  $\pm$  2.4 \\
499.5 & 94.4  $\pm$  17.9  $\pm$  6.2 & 828.3 & 118.4  $\pm$  19.9  $\pm$  7.8 & 1127.0 & 122.6 $\pm$  30.9  $\pm$  8.1 & 1368.2 & 51.2 $\pm$  26.6  $\pm$  3.4 \\
508.3 & 127.4 $\pm$  51.0  $\pm$  8.4 & 836.7 & 83.4   $\pm$  19.1  $\pm$  5.5 & 1134.2 & 65.2  $\pm$  20.9  $\pm$  4.3 & 1374.3 & 55.9 $\pm$  37.4  $\pm$  3.7 \\
517.1 & 119.9 $\pm$  27.5  $\pm$  7.9 & 845.1 & 69.7   $\pm$  17.9  $\pm$  4.6 & 1141.4 & 71.8  $\pm$  18.8  $\pm$  4.7 & 1380.6 & 22.4 $\pm$  33.5  $\pm$  1.5 \\
525.8 & 107.9 $\pm$  20.4  $\pm$  7.1 & 853.4 & 25.8   $\pm$  35.0  $\pm$  1.7 & 1148.4 & 58.0  $\pm$  17.1  $\pm$  3.8 & 1387.0 & 13.2 $\pm$  13.0  $\pm$  0.9 \\
1393.5 & 0.32  $\pm$40.2  $\pm$0.04 & \\
%
    \hhline{========}
\end{tabular} \label{tabdeut}
\end{table*}

\begin{table*}[ht]

\caption{Numerical values of  the total inclusive helicity dependent cross section
  $\Delta\sigma$ for the neutron (evaluated using  Eq.~\ref{eq:neut1}) as a function of the photon energy E$_\gamma$ in the target rest frame.
  The quoted  errors represent the 68\% confidence level of the statistical and systematic uncertainties of the experimental data, respectively. }
\begin{tabular}{cl||cl||cl||cl}
 \hhline{========}
 \textbf{E$_\gamma$} & \textbf{Value} & \textbf{E$_\gamma$} & \textbf{Value} & \textbf{E$_\gamma$} & \textbf{Value} & \textbf{E$_\gamma$} & \textbf{Value} \\
 \hline
508.3 & 118.5 $\pm$  59.7  $\pm$  9.1 & 760.5 & 70.4 $\pm$  22.9  $\pm$  13.7 & 999.2 & 15.0 $\pm$  22.5  $\pm$  10.8 & 1210.2 & 20.5 $\pm$  20.4  $\pm$  3.0 \\
517.1 & 106.5 $\pm$  37.2  $\pm$  8.7 & 769.0 & 44.8 $\pm$  26.2  $\pm$  11.8 & 1007.0 & 10.2 $\pm$  20.9  $\pm$  10.2 & 1216.8 & 17.3 $\pm$  20.8  $\pm$  2.9 \\
525.8 & 83.4 $\pm$  24.5  $\pm$  8.0 & 777.5 & 16.8 $\pm$  32.6  $\pm$  9.5 & 1014.8 & 30.8 $\pm$  20.4  $\pm$  11.5 & 1223.4 & 5.7 $\pm$  25.1  $\pm$  3.3 \\
534.6 & 97.4 $\pm$  22.5  $\pm$  9.8 & 786.0 & 52.6 $\pm$  23.7  $\pm$  9.6 & 1022.5 & 29.8 $\pm$  20.4  $\pm$  10.7 & 1229.9 & 19.9 $\pm$  29.0  $\pm$  3.0 \\
543.4 & 45.7 $\pm$  21.5  $\pm$  6.8 & 794.5 & 86.1 $\pm$  27.0  $\pm$  11.1 & 1030.2 & 24.8 $\pm$  21.1  $\pm$  10.0 & 1236.4 & 4.6 $\pm$  41.5  $\pm$  0.7 \\
552.1 & 33.5 $\pm$  23.1  $\pm$  6.3 & 803.0 & 103.2 $\pm$  25.4  $\pm$  11.2 & 1037.9 & 12.8 $\pm$  21.9  $\pm$  9.5 & 1242.9 & 29.8 $\pm$  46.8  $\pm$  2.9 \\
560.9 & 82.5 $\pm$  23.7  $\pm$  9.9 & 811.5 & 62.7 $\pm$  25.0  $\pm$  8.1 & 1045.6 & -20.9 $\pm$  21.2  $\pm$  7.3 & 1249.3 & 16.1 $\pm$  18.6  $\pm$  2.7 \\
569.6 & 40.6 $\pm$  25.6  $\pm$  8.5 & 819.9 & 43.6 $\pm$  21.6  $\pm$  6.6 & 1053.1 & 26.0 $\pm$  23.4  $\pm$  8.9 & 1255.6 & 66.4 $\pm$  31.3  $\pm$  5.1 \\
578.4 & 126.5 $\pm$  29.2  $\pm$  13.7 & 828.3 & 95.7 $\pm$  23.4  $\pm$  8.7 & 1060.7 & 67.7 $\pm$  24.6  $\pm$  10.3 & 1261.9 & 40.4 $\pm$  27.7  $\pm$  3.4 \\
587.1 & 67.3 $\pm$  42.8  $\pm$  10.0 & 836.7 & 63.4 $\pm$  22.5  $\pm$  6.2 & 1068.3 & 54.6 $\pm$  20.5  $\pm$  8.5 & 1268.2 & 27.2 $\pm$  23.3  $\pm$  2.3 \\
595.9 & 109.0 $\pm$  25.2  $\pm$  13.0 & 845.1 & 49.7 $\pm$  21.6  $\pm$  5.2 & 1077.6 & 46.3 $\pm$  30.5  $\pm$  7.1 & 1274.4 & 18.1 $\pm$  32.6  $\pm$  2.8 \\
604.6 & 77.8 $\pm$  34.7  $\pm$  11.6 & 853.4 & 8.9 $\pm$  39.2  $\pm$  2.2 & 1083.2 & 26.0 $\pm$  21.4  $\pm$  5.4 & 1280.6 & 26.3 $\pm$  32.6  $\pm$  4.7 \\
613.4 & 114.3 $\pm$  35.4  $\pm$  14.5 & 861.7 & 43.7 $\pm$  29.4  $\pm$  3.8 & 1090.6 & -3.5 $\pm$  20.6  $\pm$  4.4 & 1286.7 & 11.1 $\pm$  44.4  $\pm$  2.5 \\
622.1 & 59.4 $\pm$  25.3  $\pm$  12.4 & 870.0 & 69.6 $\pm$  27.4  $\pm$  5.8 & 1098.0 & 23.6 $\pm$  21.2  $\pm$  5.8 & 1292.8 & 10.4 $\pm$  39.5  $\pm$  1.8 \\
630.8 & 140.9 $\pm$  44.4  $\pm$  18.5 & 878.3 & 6.7 $\pm$  22.5  $\pm$  3.6 & 1105.3 & 39.3 $\pm$  23.8  $\pm$  5.3 & 1298.8 & -13.7 $\pm$  59.2  $\pm$  0.8 \\
641.7 & 107.9 $\pm$  43.1  $\pm$  17.8 & 886.5 & 49.7 $\pm$  22.7  $\pm$  6.1 & 1112.6 & 67.1 $\pm$  22.4  $\pm$  6.6 & 1304.8 & 61.8 $\pm$  35.0  $\pm$  4.9 \\
648.2 & 106.2 $\pm$  25.4  $\pm$  17.8 & 894.8 & 12.0 $\pm$  28.0  $\pm$  3.7 & 1119.9 & 39.2 $\pm$  28.2  $\pm$  5.1 & 1310.7 & 10.7 $\pm$  27.7  $\pm$  2.6 \\
656.9 & 149.0 $\pm$  29.3  $\pm$  20.6 & 903.0 & 28.6 $\pm$  21.1  $\pm$  5.7 & 1127.0 & 103.4 $\pm$  36.5  $\pm$  8.9 & 1316.6 & 16.1 $\pm$  29.6  $\pm$  2.5 \\
665.6 & 107.1 $\pm$  25.4  $\pm$  19.2 & 911.2 & 34.3 $\pm$  20.7  $\pm$  6.0 & 1134.2 & 49.6 $\pm$  26.6  $\pm$  4.8 & 1322.5 & -11.0 $\pm$  27.7  $\pm$  1.2 \\
674.3 & 52.8 $\pm$  25.6  $\pm$  16.8 & 919.3 & 69.2 $\pm$  23.4  $\pm$  7.7 & 1141.4 & 51.4 $\pm$  24.2  $\pm$  5.4 & 1328.3 & 32.1 $\pm$  36.3  $\pm$  3.7 \\
683.0 & 67.5 $\pm$  25.1  $\pm$  18.1 & 927.4 & 42.0 $\pm$  23.6  $\pm$  7.1 & 1148.4 & 30.0 $\pm$  23.1  $\pm$  4.6 & 1334.0 & 32.5 $\pm$  48.0  $\pm$  4.2 \\
691.6 & 110.4 $\pm$  26.0  $\pm$  20.3 & 935.5 & 51.2 $\pm$  27.1  $\pm$  9.3 & 1155.5 & 7.1 $\pm$  25.0  $\pm$  3.3 & 1339.7 & 51.8 $\pm$  46.5  $\pm$  5.7 \\
700.3 & 120.2 $\pm$  29.1  $\pm$  20.0 & 943.6 & 32.1 $\pm$  22.4  $\pm$  8.0 & 1162.4 & 59.5 $\pm$  20.7  $\pm$  6.1 & 1345.4 & 53.2 $\pm$  41.9  $\pm$  4.5 \\
708.9 & 80.2 $\pm$  36.2  $\pm$  18.0 & 951.6 & 23.4 $\pm$  24.6  $\pm$  7.8 & 1169.4 & 38.2 $\pm$  23.5  $\pm$  4.4 & 1351.0 & 36.4 $\pm$  41.5  $\pm$  2.4 \\
717.6 & 120.1 $\pm$  28.3  $\pm$  19.6 & 959.6 & 42.4 $\pm$  21.8  $\pm$  10.4 & 1176.3 & 26.0 $\pm$  21.8  $\pm$  3.0 & 1356.6 & 57.7 $\pm$  50.3  $\pm$  5.1 \\
726.2 & 132.3 $\pm$  26.8  $\pm$  19.8 & 967.6 & 24.4 $\pm$  25.9  $\pm$  8.1 & 1183.2 & 45.0 $\pm$  23.5  $\pm$  3.5 & 1362.3 & 9.3 $\pm$  37.6  $\pm$  3.2 \\
734.8 & 21.1 $\pm$  29.5  $\pm$  14.7 & 975.5 & 20.9 $\pm$  22.7  $\pm$  10.0 & 1190.0 & 67.5 $\pm$  25.3  $\pm$  4.6 & 1368.2 & 37.1 $\pm$  33.5  $\pm$  3.8 \\
743.3 & 95.2 $\pm$  28.1  $\pm$  18.3 & 983.5 & 38.0 $\pm$  20.7  $\pm$  9.1 & 1196.8 & 46.2 $\pm$  23.3  $\pm$  2.9 & 1374.3 & 40.2 $\pm$  44.7  $\pm$  4.2 \\
751.9 & 63.2 $\pm$  27.9  $\pm$  15.2 & 991.3 & 13.5 $\pm$  23.2  $\pm$  8.8 & 1203.5 & 40.3 $\pm$  21.3  $\pm$  3.3 & 1380.6 & -0.5 $\pm$  38.4  $\pm$  2.3 \\
1387.0 & -6.2 $\pm$ 51.4 $\pm 1.6 $ & 1393.0  & -25.9 $\pm$  72.1 $\pm$ 1.7 & & \\
%
%
    \hhline{========}
\end{tabular} \label{tabneut}
\end{table*}

\clearpage
 
%
%
%
%

%
\end{document}